\documentclass[11pt]{article}
\usepackage{cite}
\usepackage{amsmath,amscd,amsfonts,mathtools,amssymb}
\usepackage[small,bf,hang]{caption}
\usepackage{slashed}
\usepackage{mathabx}
\usepackage{latexsym,epsfig}
\usepackage{stmaryrd}

\usepackage[vcentermath]{youngtab}

\usepackage[usenames,dvipsnames]{color}
\usepackage{amssymb}
\usepackage{breqn}

\def\hybrid{
        \topmargin -20pt
        \oddsidemargin 0pt
        \headheight 0pt \headsep 0pt
        \textwidth 6.25in 
        \textheight 9.5in 
        \marginparwidth .875in
        \parskip 5pt plus 1pt \jot = 1.5ex}

\hybrid

 \csname
@addtoreset\endcsname{equation}{section}

\def\moth{\mathsurround=0pt}
\newdimen\zo \zo=0pt

\def\tick{\leaders\hrule height 0.5ex depth 0pt \hskip 0.5pt}
\def\upboxfill{$\moth \setbox\zo\hbox{\tick}%
  \hskip 3pt\hbox to 0pt{$\tick$\hss}\hrulefill \hbox to 7.5pt{$\tick$\hss}$}

\def\dtick{\leaders\hrule height .34pt depth 0.5ex \hskip 0.5pt}
\def\downboxfill{$\moth \setbox\zo\hbox{\dtick}%
  \hskip 2pt\hbox to 0pt{$\dtick$\hss}\hrulefill \hbox to 2pt{$\dtick$\hss}$}

\def\bec{\begin{center}}
\def\ec{\end{center}}

\def\nn{\nonumber}

\def\be{\begin{equation}}
\def\ee{\end{equation}}
\newcommand{\beq}{\begin{equation}\begin{aligned}}
\newcommand{\eeq}{\end{aligned}\end{equation}}
\def\bea{\begin{eqnarray}}
\def\eea{\end{eqnarray}}
\def\ba{\begin{array}}
\def\ea{\end{array}}

\allowdisplaybreaks[2]

\begin{document}

\begin{titlepage}

\begin{center}
\vskip 1.6cm
{\Large \bf {The generalized Bergshoeff-de Roo identification}}\\
 \vskip 2.0cm
{\large {W. Baron$^{1}$, E. Lescano$^2$ and D. Marques$^2$}}
\vskip 0.5cm

$^1$ {\it Instituto de F\'isica La Plata (CONICET-UNLP)}\\ {\it Calle 49 y 115, La Plata, Argentina} \\[1ex]

$^2$ {\it Instituto de Astronom\'ia y F\'isica del Espacio (CONICET-UBA)}\\ {\it Ciudad Universitaria, Buenos Aires, Argentina} \\[1ex]

\vskip 0.5cm

{\small \verb"wbaron@fisica.unlp.edu.ar , {elescano, diegomarques}@iafe.uba.ar"}

\vskip 1cm
{\bf Abstract}	
\end{center}

\vskip 0.2cm

\noindent
\begin{narrower}

{\small
There are two main approaches to duality covariant first order higher derivative corrections to the heterotic string, one extending the duality structure and the other deforming the gauge transformations. In this paper we introduce a framework from which both approaches can be derived, proving their equivalence and extending them to higher orders.}

\end{narrower}

\vskip 1.5cm

\end{titlepage}

\tableofcontents

\vspace{0.5cm}

\section{Introduction}
T-duality is an exact symmetry of (super)string theories. It leaves hidden imprints in the low-energy supergravity limits, which can be unmasked through toroidal compactifications. They are however already present even before dimensional reduction, as realized from the advent of duality symmetric approaches.
Double Field Theory (DFT) \cite{Siegel:1993th},\cite{Hull:2009mi} (for reviews see \cite{Aldazabal:2013sca}) is a framework that makes the T-duality symmetry manifest, thus highly constraining the allowed interactions of the effective theories under its reach. There are many variants and extensions, but the ones that are of interest here are the frame-like \cite{Hohm:2010xe} or flux \cite{Geissbuhler:2013uka} formulations of the $D=10$ and ${\cal N} = 1$ supersymmetric \cite{Hohm:2011nu},\cite{Jeon:2011sq} heterotic DFT \cite{Hohm:2011ex}.

Higher derivative corrections to (super)gravity arising from string theory also share these hidden imprints. Toroidal compactifications can again make them appear in the form of a symmetry in the lower dimensional theory \cite{Meissner:1996sa}, though it turns out that  the problem is a little more involved than the two-derivative case because unconventional non-covariant field redefinitions are required.  Many methods have been proposed to learn about higher derivative interactions based on the way in which duality organizes the lower dimensional theories \cite{methods}.

The natural scenario to study how T-duality constrains higher derivative interactions is DFT, as it encodes duality symmetries from the onset in a background independent fashion, so it is not surprising that this long standing problem  has resurfaced in such a context \cite{Hohm:2013jaa}.  Interestingly, two seemingly different approaches arise in this scenario for the case of the heterotic string to first order in $\alpha'$:
\begin{itemize}
\item One scenario considers a generalized version of the Green-Schwarz transformation. In the standard picture \cite{Green:1984sg} the Kalb-Ramond field is not Lorentz invariant, which requires Lorentz Chern-Simons corrections to its three-form field strength $\widehat H$ with respect to a certain torsionful spin connection $\omega_- = \omega - \frac 1 2 \widehat H$, due to Hull \cite{Hull:1986kz} (recently discussed in \cite{delaOssa:2014msa}). The first order T-duality covariant generalization was introduced in \cite{Marques:2015vua}, \cite{Baron:2017dvb}. There, the fields are $O(D,D)$ multiplets transforming as usual under generalized diffeomorphisms but receiving a first order generalized Green-Schwarz transformation that deforms the double Lorentz symmetry. After a $GL(D)$ decomposition, choice of solution to the strong constraint and proper field redefinitions the standard Green-Schwarz transformation is recovered from the generalized one. Due to the T-duality covariance the Kalb-Ramond and gravitational sectors mix in such a way that the generalized Green-Schwarz transformation demands the well-known quadratic Riemann interactions, on top of the Chern-Simons terms.

\item Another scenario is based on the work of Bergshoeff and de Roo \cite{Bergshoeff:1988nn},\cite{Bergshoeff:1989de}. There, the torsionful spin connection $\omega_-$ and the gravitino curvature were shown to behave effectively as a gauge multiplet with respect to supersymmetry to first order in $\alpha'$. We refer to this as {\it the Bergshoeff-de Roo identification} between independent and composite degrees of freedom. This fact was exploited to compute the first order corrections to the action \cite{Bergshoeff:1988nn}, and later extended up to quartic Riemann interactions through a Noether procedure \cite{Bergshoeff:1989de}. In \cite{Bedoya:2014pma}, \cite{Coimbra:2014qaa} this idea was engineered by considering the extended duality structure $O(D,D+k)$ of the heterotic setup \cite{Hohm:2011ex}, such that after a $GL(D)$ decomposition the one-form gauge fields were identified with (a component of) the generalized spin connection for $O(1,D-1) \in O(1, D+k - 1)$, extensively discussed in the literature \cite{Siegel:1993th},\cite{Hohm:2010xe},\cite{Coimbra:2011nw},\cite{Jeon:2011cn},\cite{Hohm:2011si}. Since the identification between independent and composite degrees of freedom is done {\it after} the $GL(D)$ decomposition, duality covariance is not manifest and must be checked explicitly.
\end{itemize}
Both approaches look rather orthogonal as in the first case the duality structure is extended  and the local symmetries remain intact, whereas in the second case the duality group stays unmodified but the local symmetries are deformed. The link between them remains unclear -though they lead to the same first order heterotic action- and higher order corrections to these approaches have been so far elusive. In this paper we introduce a general setup from which both approaches can be derived and extended to all orders in a derivative expansion. Let us provide some highlights of the route we follow and the results we encounter:
\begin{itemize}
\item We consider a hybrid between both approaches. Starting from a ${\cal G} = O(D,D+k)$ extended space, we perform a $G = O(D,D)$ decomposition -as opposed to the standard $GL(D)$ decomposition- in the lines of \cite{HSZheterotic}. This leaves us with a $G$-valued generalized frame plus additional covariantly constrained $G$-vectors, thus fully preserving T-duality covariance.

\item The ${\cal G}$ generalized diffeomorphisms induce a {\it generalized Green-Schwarz transformation} for the $G$-valued frame with respect to the extra $G$-vectors.

\item The $G$-vectors are identified in a duality covariant way -by matching gauge and supersymmetry transformations- with generalized fluxes playing the role of a generalized spin connection for the {\it full} $O(1, D+k-1)$. We do the same between the gauginos and a generalized notion of gravitino curvature. This is what we call the {\it generalized Berghshoeff-de Roo identification}.
\end{itemize}

Contrary to the standard identification, the generalized one is exact. This is possible because the gauge group is identified with the full $O(1,D+k-1)$ rather than its $O(1,D-1)$ subgroup, so the generalized identification that we propose here would be impossible to implement in supergravity. An interesting consequence of our results is that the tangent space must be infinite dimensional. This might sound strange, but is somewhat expected given that the identification is exact, which presumably means that it captures an infinite tower of higher derivative corrections. We test this proposal by performing an $\alpha'$ perturbative expansion. To first order we recover the well known heterotic generalized Green-Schwarz transformation of \cite{Marques:2015vua}, and to second order we obtain a novel consistent correction that preserves the constraints and closes.

The paper is organized as follows. In Section \ref{sec::extended} we introduce the ${\cal G}$ structure of the extended space and reduce it to the double space in a $G$-covariant way. Section \ref{sec::gGG} is devoted to lock the vector degrees of freedom in terms of  (derivatives of) the $G$-covariant generalized frame (the generalized Bergshoeff-de Roo identification). As a consistency check of the approach we perform a derivative expansion, finding the second order completion of the heterotic generalized Green-Schwarz transformation of \cite{Marques:2015vua}. The full supersymmetric treatment is presented in Section \ref{sec::susy}. An outlook is given in the last section, together with a list of possible future lines of research.

\section{From extended to double} \label{sec::extended}

The starting point of this section is the generalized frame formalism \cite{Hohm:2010xe} (mostly following the conventions in \cite{Geissbuhler:2013uka}) in the
\be
{\rm Extended \ space}\ \ \ \ \ \ \ \left\{\begin{matrix}{\rm Global\ symmetry}\ \ \ {\cal G} = O(D,D+k) \ \ \ \ \ \  \ \ \ \ \ \  \ \ \ \ \ \ \ \ \ \ \\ {\rm Local \ symmetry} \ \ \ {\cal H} = \underline{O(D-1,1)} \times \overline{O(1,D+k-1)} \end{matrix} \right. \ \  ,
\ee
and the goal is to re-formulate it in terms of multiplets of the
\be
{\rm Double \ space}\ \ \ \ \ \ \ \ \ \left\{\begin{matrix}{\rm Global\ symmetry}\ \ \ G = O(D,D) \ \ \ \ \ \  \ \ \ \ \ \  \ \ \ \ \ \ \ \ \ \ \\ {\rm Local \ symmetry} \ \ \ H = \underline{O(D-1,1)} \times \overline{O(1,D-1)}\end{matrix} \right.\ \ . \ \ \ \
\ee

\begin{table}[ht]
\begin{center}
\begin{tabular}{|l|l|l|l|} \hline
{\bf Name}& {\bf Group}  & {\bf Indices}  &  {\bf Metric} \\ \hline
$G$ $\begin{matrix} \\ ~\end{matrix}$  & $O(D,D)$ & $M$ & $\eta_{M N}$ \\ \hline
$g$ $\begin{matrix} \\ ~\end{matrix}$  & $O(k)$ & $\alpha$ & $\kappa_{\alpha \beta}$ \\ \hline
$\cal G$ & $O(D,D+k)$ & ${\cal M} = (M,\,\alpha)$ &
$\eta_{\cal M N} = \left(\begin{matrix}\eta_{M N} & 0 \\ 0 & \kappa_{\alpha \beta}\end{matrix}\right)$ \\ \hline
$\underline {\cal H}$ = $\underline H$ $\begin{matrix} \\ ~\end{matrix}$ & $\underline{O(D-1,1)}$ & $\underline{\cal A} = \underline a$ & $P_{\underline {\cal  A} \underline {\cal B}} = P_{\underline a \underline b}$ \\ \hline
$\overline H$ $\begin{matrix} \\ ~\end{matrix}$  & $\overline{O(1,D-1)}$ & $\overline a$ & $\bar P_{\overline {a b}}$ \\ \hline
$\overline h$ $\begin{matrix} \\ ~\end{matrix}$ & $\overline{O(k)}$ & $\overline \alpha$ & $\kappa_{\overline {\alpha \beta}}$ \\ \hline
$\overline {\cal H}$ & $\overline{O(1,D+k-1)}$ & $\overline{\cal A} = (\overline a,\, \overline \alpha)$ & $\bar P_{\overline {\cal A} \overline {\cal B}} = \left(\begin{matrix} \bar P_{\overline {a b}} & 0 \\ 0 & \kappa_{\overline {\alpha \beta}}\end{matrix}\right)$ \\ \hline
$\cal H$ & $\underline{{\cal H}} \times \overline{{\cal H}}$ & ${\cal A} = (\underline{{\cal A}},\,\overline{{\cal A}}) = (\underline a,\, \overline a ,\, \overline \alpha)$ & $\eta_{\cal A B} = \left(\begin{matrix}P_{\underline {\cal A B}} & 0 \\ 0 & \bar P_{\overline {\cal A B}} \end{matrix} \right)$  \\ \hline
$H$ & $\underline{H}\times \overline{H}$ & $A = (\underline a,\, \overline a)$ & $\eta_{A B} = \left(\begin{matrix}P_{\underline a \underline b}& 0 \\ 0 & \bar P_{\overline {ab}}\end{matrix}\right)$\\ \hline
\end{tabular}
    \end{center}
    \caption{Notation used throughout the paper. Modulo a few exceptions, calligraphic letters refer to the extended space while conventional ones to the double space. The metrics and their inverses are used to raise and lower indices.} \label{Notation}
\end{table}

The notation is presented in Table  \ref{Notation}. The degrees of freedom in the extended case are a generalized dilaton $d$ and a generalized frame ${\cal E}_{\cal M}{}^{\cal A}$ which is a constrained field, satisfying
\be
\eta_{\cal M N} = {\cal E}_{\cal M}{}^{{\cal A}}\, \eta_{\cal A B}\, {\cal E}_{{\cal N}}{}^{{\cal B}}\ . \label{Constraint}
\ee
The generalized dilaton plays no role whatsoever in our analysis, so it will be ignored. The gauge transformations of the extended generalized frame are given in terms of extended and gauged generalized diffeomorphisms, plus extended Lorentz ${\cal H}$-transformations
\bea
\delta {\cal E}_{\cal M}{}^{\cal A} = \xi^{\cal P} \partial_{\cal P} {\cal E}_{\cal M}{}^{\cal A} + \left(\partial_{\cal M} \xi^{\cal P} - \partial^{\cal P} \xi_{\cal M}\right) {\cal E}_{\cal P}{}^{\cal A} + g \, f_{\cal M N}{}^{\cal P} \xi^{\cal N} {\cal E}_{\cal P}{}^{\cal A} + {\cal E}_{\cal M}{}^{\cal B}\, \Gamma_{\cal B}{}^{\cal A} \ . \label{TransformationExtended}
\eea
We included a dimensional coupling $g^{-2} \sim \alpha'$ to render the structure constants $f_{\cal M N P}$ dimensionless.  The extended Lorentz parameters satisfy
\bea
\Gamma_{\cal A B} = \Gamma_{[{\cal A B}]} \ , \ \ \ \ \Gamma_{\underline {\cal A} \overline {\cal B}} = 0\ ,
\eea
and the gaugings obey linear and quadratic constraints
\be
f_{\cal M N P} = f_{[{\cal M N P}]} \ , \ \ \ \ f_{[{\cal M N}}{}^{\cal K} f_{{\cal P}]{\cal K}}{}^{\cal L} = 0\ . \label{LinearQuadratic}
\ee

The gauge transformations close
\be
\left[\delta_1 ,\, \delta_2\right] = - \delta_{12} \ ,
\ee
with respect to the following brackets
\bea
\xi^{\cal M}_{12} &=& 2 \xi_{[1}^{\cal P} \partial_{\cal P} \xi_{2]}^{\cal M} + \partial^{\cal M} \xi_{[1}^{\cal P} \xi_{2]{\cal P}} + g \, f_{\cal N P}{}^{\cal M} \xi_1^{\cal N}  \xi_2^{\cal P} \label{ExtendedBracketXi}\\
\Gamma_{12 {\cal A B}} &=& 2 \xi_{[1}^{\cal P} \partial_{\cal P} \Gamma_{2]{\cal A B}} + \Gamma_{[1 {\cal A}}{}^{\cal C} \, \Gamma_{2] {\cal B C}}  \ , \label{ExtendedBracketGamma}
\eea
provided a strong constraint is imposed
\be
\eta^{\cal M N}\, \partial_{\cal M} \otimes \partial_{\cal N} = 0 \ , \ \ \ \ \ \ f_{\cal M N}{}^{\cal P} \, \partial_{\cal P} = 0 \ . \label{ExtendedStrongConstraint}
\ee

We now want to write everything in terms of $G = O(D,D) \in {\cal G}$ multiplets. Taking a close look into Table \ref{Notation} we see that derivatives and parameters split as follows
\be
\partial_{\cal M} = (\partial_M,\, \partial_\alpha) \ , \ \ \ \  \xi^{\cal M} = \left(\xi^M,\, \xi^\alpha\right) \ .
\ee
Since we intend to preserve $G$-invariance, we need to annihilate all the $G$ components of the gaugings $f_{\cal M N P}$, such that only the $f_{\alpha \beta \gamma}$ survive. The extended strong constraint \eqref{ExtendedStrongConstraint} then implies the double strong constraint
\be
\partial_\alpha = 0 \ , \ \ \ \ \eta^{M N} \, \partial_M \otimes \partial_N = 0 \ .
\ee
As for the parameters, the $\xi^M$ components now generate double generalized diffeomorphisms, and the $\xi^\alpha$ generate gauge transformations of a given gauge group $\cal K$ defined by its structure constants $f_{\alpha\beta\gamma}$.

The extended generalized frame satisfies the constraint \eqref{Constraint} implying that it contains $dim[{\cal G}]$ degrees of freedom (dof)
\be
dim[O(D,D+k)] = \underbrace{dim[O(D,D)]}_{E_M{}^A} + \underbrace{2 D k}_{A_M{}^\alpha} + \underbrace{dim[O(k)]}_{e_\alpha{}^{\overline \alpha}} \ ,
\ee
where $E_M{}^A$ is a double generalized frame satisfying
\be
\eta_{M N} = E_M{}^{A} \eta_{A B} E_{N}{}^B \ ,
\ee
and $e_\alpha{}^{\overline \alpha}$ is a bijective map between $\overline h$ and $g$
\be
\kappa_{\alpha \beta} = e_\alpha{}^{\overline \alpha} \, \kappa_{\overline{\alpha \beta}} \, e_\beta{}^{\overline \beta} \ . \label{bijection}
\ee
Then, the extended dof admit a general $G$ and $H$ decomposition of the form
\bea
{\cal E}_M{}^A &=& (\chi^{\frac 1 2 }){}_M{}^N\, E_N{}^A\ , \nn\\
{\cal E}_M{}^{\overline \alpha} &=& - {\cal A}_M{}^\beta \, e_\beta{}^{\overline \alpha} \ ,\label{FrameParameterizationA}\\
{\cal E}_\alpha{}^A &=& {\cal A}^M{}_\alpha\, E_M{}^A\ , \nn\\
{\cal E}_\alpha{}^{\overline \alpha} &=& (\Box^{\frac 1 2}){}_\alpha{}^\beta \, e_\beta{}^{\overline \alpha}\ , \nn
\eea
where we introduced the following quantities
\bea
\chi_{M N} = \eta_{M N} - {\cal A}_M{}^\alpha\, {\cal A}_{N\alpha} \ , \ \ \ \ \ \Box_{\alpha \beta} = \kappa_{\alpha \beta} - {\cal A}_{M \alpha} \, {\cal A}^M{}_\beta \ ,
\eea
that satisfy the identity
\be
{\cal A}_M{}^\beta\, f(\Box){}_\beta{}^\alpha = f(\chi){}_M{}^N \, {\cal A}_N{}^\alpha \ ,
\ee
for any function $f$.

Due to the original $\cal H$ symmetry, there are many non-physical gauge dof. It will then turn out to be convenient to perform a gauge fixing to remove some of them
\be
E^{M \overline a}\, {\cal A}_{M}{}^\alpha = 0\ ,   \ \ \ \ \ e_\alpha{}^{\overline  \alpha} = constant\ .
\ee
Demanding that these constraints are gauge invariant $\delta {\cal E}_\alpha{}^{\overline a} = 0$ and $\delta e_\alpha{}^{\overline \alpha} = 0$  freezes the following components of the $\overline{{\cal H}}$ parameters
\bea
\Gamma_{\overline {\alpha a}} &=& e^\alpha{}_{\overline \alpha} \, (\Box^{- \frac 1 2}){}_\alpha{}^\beta\, \partial_P \xi_\beta \, E^P{}_{\bar a}\ , \label{GaugedFixedParameters}\\
\Gamma_{\overline {\alpha \beta}} &=&  e^\alpha{}_{[\overline \alpha} \, e^\beta{}_{\overline \beta]} \, (\Box^{-\frac 1 2}){}_\alpha{}^\gamma \left(\delta (\Box^{\frac 1 2}){}_{\gamma \beta} - \partial^P \xi_\gamma\, {\cal A}_{P \beta} - g\, f_{\gamma \delta}{}^\lambda \, \xi^\delta \,(\Box^{\frac 1 2}){}_{\lambda \beta} \right)
\ , \nn
\eea
forcing a dependence on the generators of $\cal K$.

 We now compute from the gauge transformations in the extended space \eqref{TransformationExtended} how the double dof in \eqref{FrameParameterizationA} transform. The transformations turn out to have simpler expressions in terms of a redefined vector field ${\cal C}_M{}^\alpha$
\be
{\cal C}_M{}^\alpha = - {\cal A}_{M}{}^\beta (\Box^{-\frac 1 2})_\beta{}^\alpha \ , \ \ \ \ {\cal A}_M{}^\alpha = - {\cal C}_M{}^\beta (\Delta^{-\frac 1 2})_\beta{}^\alpha \ , \label{RedefinitionVectors}
\ee
which is also covariantly constrained $E^{M \overline a} {\cal C}_M{}^\alpha = 0$, and we define
\be
\Delta_{\alpha \beta} = \kappa_{\alpha \beta} + {\cal C}_{M \alpha} {\cal C}^M{}_\beta \ , \ \ \ \ \Xi_{M N} = \eta_{M N}  + {\cal C}_M{}^\alpha {\cal C}_{N\alpha} \ ,
\ee
which relate to the previous definitions as follows
\bea
\Delta_{\alpha \beta} = (\Box^{-1}){}_{\alpha \beta} \ , \ \ \ \ \chi_{M N} = (\Xi^{-1}){}_{M N} \ , \ \ \ \ {\cal C}_M{}^\beta f(\Delta){}_\beta{}^\alpha = f(\Xi){}_M{}^N {\cal C}_N{}^\alpha \ .
\eea
The transformations of the double fields are
\bea
\delta E_M{}^{\overline a} &=& \widehat {\cal L}_\xi E_M{}^{\overline a} + E_M{}^{\overline b}\, \Lambda_{\overline b}{}^{\overline a} + E^{P \overline a}\, \partial_P \xi^\alpha \, {\cal C}_{M \alpha} \ , \label{TransformationEoverlineA}\\
\delta E_M{}^{\underline a} &=& \widehat {\cal L}_\xi E_M{}^{\underline a} + E_M{}^{\underline b}\, \Lambda_{\underline b}{}^{\underline a} - \partial_{\overline M} \xi^\alpha \, {\cal C}_{Q \alpha} \, E^{Q \underline a} \ , \label{TransformationEunderlineA}\\
\delta {\cal C}_M{}^\alpha &=& \widehat {\cal L}_\xi {\cal C}_M{}^\alpha + \partial_{\underline M} \xi^\alpha - \partial_{\overline M} \xi^\beta {\cal C}_{Q \beta} {\cal C}^{Q \alpha} + {\cal C}_M{}^\beta\, \partial^P \xi_\beta\, {\cal C}_P{}^\alpha + g\, f_{\beta \gamma}{}^\alpha \xi^\beta {\cal C}_M{}^\gamma \ , \ \ \ \label{TransformationC}
\eea
where $\widehat {\cal L}$ denotes the standard generalized Lie derivative in the double space, and we have redefined the double Lorentz parameters so that both projections of the frame field look slightly more symmetric
\bea
\Lambda_{\underline {a b}} &=& \Gamma_{\underline {a b}} - E^M{}_{[\underline a} E^N{}_{\underline b]} (\Xi^{\frac 1 2}){}_M{}^P \left(\delta (\Xi^{-\frac 1 2})_{P N} + \partial_P \xi^\alpha {\cal C}_{Q \alpha} (\Xi^{-\frac 1 2}){}^Q{}_N\right)\ , \label{LorentzParamterRedefinition}\\
\Lambda_{\overline {a  b}} &=& \Gamma_{\overline {a b}}  \ . \nn
\eea
In addition we defined the standard projected derivatives
\be
\partial_{\underline M} = P_M{}^N \partial_N \ , \ \ \ \ \partial_{\overline M} = \bar P_M{}^N \partial_N \ ,
\ee
in terms of the projectors
\be
P_{M N} = E_M{}^{\underline a} E_{N \underline a} = \frac 1 2 \left(\eta_{M N} - {\cal H}_{M N}\right) \ , \ \ \ \ \bar P_{M N} = E_M{}^{\overline a} E_{N \overline a} = \frac 1 2 \left(\eta_{M N} + {\cal H}_{M N}\right) \ .
\ee

In terms of the collective $H$  indices of Table \ref{Notation} the transformation of the frame field can be recast as
\be
\delta E_M{}^A = \widehat {\cal L}_\xi E_M{}^A + E_M{}^B \Lambda_B{}^A - 2 \partial_{[\overline M} \xi^\alpha \, {\cal C}_{\underline N] \alpha}\, E^{N A}\ ,
\ee
and this reproduces the schematic form of generalized Green-Schwarz transformations discussed in \cite{Marques:2015vua} for a given choice in the bi-parametric freedom discussed there. The difference is that here ${\cal C}_M{}^\alpha$ is an independent degree of freedom, corresponding to the duality covariant gauge vectors of a gauge group $\cal K$. Normally, in the heterotic supergravity ${\cal C}_M{}^\alpha$ would undergo a $GL(D)$ parameterization in terms of one-form gauge fields $A_\mu{}^\alpha$ and the gauge group would be ${\cal K} = SO(32)$ or ${\cal K} = E_8 \times E_8$, or further enhancements \cite{Aldazabal:2017jhp},\cite{Aldazabal:2018uzm},\cite{Fraiman:2018ebo} (see also \cite{Cho:2018alk} for an alternative approach to heterotic DFT). In this paper we will ignore the gauge sector of heterotic supergravity, so we have different plans for ${\cal C}_M{}^\alpha$.

We have then extracted the double field transformations from those in the extended space. Closure is guaranteed by construction, and the brackets can be obtained either through direct inspection or from the $G$ and $H$ decomposition of the extended brackets \eqref{ExtendedBracketXi}-\eqref{ExtendedBracketGamma}. In the latter case, the field-dependent redefinition of the $\underline{\cal H}$ parameter \eqref{LorentzParamterRedefinition} must be properly accounted for. The double brackets are
\bea
\xi_{12}^M &=& 2 \xi_{[1}^{P} \partial_{P} \xi_{2]}^{M} + \partial^{M} \xi_{[1}^{P} \xi_{2]{P}}  + \partial^M \xi_{[1}^\alpha \xi_{2]\alpha} \ , \nn \\
\xi_{12}^\alpha &=& 2 \xi_{[1}^P \partial_P \xi_{2]}^\alpha + g f_{\beta \gamma}{}^\alpha \xi_1^\beta \xi_2^\gamma \ , \label{extendedBrackets} \\
\Lambda_{12 \overline {a b}} &=& 2 \xi_{[1}^P \partial_P \Lambda_{2] \overline {a b}} + 2 \Lambda_{[1 \overline a}{}^{\overline c} \Lambda_{2] \overline {bc}}  + 2 E^M{}_{[\overline a} E^N{}_{\overline b]}\partial_{M} \xi_1^\alpha \partial_{N} \xi_2^\beta \, \Delta_{\alpha \beta} \ , \nn\\
\Lambda_{12 \underline {a  b}} &=& 2 \xi_{[1}^P \partial_P \Lambda_{2] \underline {a b}} + 2 \Lambda_{[1 \underline a}{}^{\underline c} \Lambda_{2] \underline {bc}}  + {\cal C}_{[\underline a}{}^\alpha {\cal C}_{\underline b]}{}^\beta \, {\cal H}^{M N} \partial_M \xi_{1 \alpha} \partial_N \xi_{2 \beta} + 2 E^M{}_{[\underline a} E^N{}_{\underline b]} \partial_{M} \xi_1^\alpha \partial_{N} \xi_{2 \alpha} \ . \nn
\eea

~

We include here a few words on the generalized metric formulation, mostly intended to show that the $O(D,D)$ decomposition we use is exactly the one introduced in \cite{HSZheterotic}. The extended generalized metric is defined as follows
\bea
\widehat {\cal H}_{\cal M N} = {\cal E}_{\cal M}{}^{\overline {\cal A}}{\cal E}_{{\cal N} \overline {\cal A}} - {\cal E}_{\cal M}{}^{\underline {\cal A}}{\cal E}_{{\cal N} \underline {\cal A}} = \left(\begin{matrix}\widetilde {\cal H}_{M N} & \widetilde {\cal C}_{M \beta} \\ \widetilde {\cal C}_{N\alpha} & \widetilde {\cal N}_{\alpha \beta} \end{matrix}\right) \ .
\eea
The components can be computed directly from \eqref{FrameParameterizationA} and after implementing the redefinition (\ref{RedefinitionVectors}) we find
\bea
\widetilde {\cal H}_{M N} &=& {\cal H}_{M N} + 2 {\cal C}_M{}^\alpha (\Delta^{-1}){}_{\alpha \beta} {\cal C}_N{}^\beta \ , \nn \\
\widetilde {\cal C}_{M \alpha} &=&  2 {\cal C}_{M}{}^\beta (\Delta^{-1}){}_{\beta \alpha} \ ,\\
\widetilde {\cal N}_{\alpha \beta} &=& - \kappa_{\alpha \beta} + 2 (\Delta^{-1}){}_{\alpha \beta} \ . \nn
 \eea
This is precisely the parameterization of the extended space generalized metric in terms of $O(D,D)$ multiplets as presented in \cite{HSZheterotic}. We are using the same letters, but strictly the tensors in \cite{HSZheterotic} contain scalars in the context of heterotic compactifications on tori, while here we are dealing with the full generalized fields and not assuming a compactification. For completeness we give the transformation of the double generalized metric
\be
\delta {\cal H}_{M N} = \widehat {\cal L}_\xi {\cal H}_{M N} + 4 \partial_{(\overline M} \xi^\alpha \, {\cal C}_{\underline N)\alpha} \ ,
\ee
which again takes the form of a generalized Green-Schwarz transformation as in \cite{Marques:2015vua}.

~

Key to the forthcoming analysis are the extended generalized fluxes
\bea
{\cal F}_{\cal A B C} = 3 \, {\cal D}_{[{\cal A}} {\cal E}^{\cal M}{}_{\cal B} \, {\cal E}^{\cal N}{}_{{\cal C}]} \, \eta_{\cal M N} + g \, f_{\cal M N P}\, {\cal E}^{\cal M}{}_{\cal A}{\cal E}^{\cal N}{}_{\cal B}{\cal E}^{\cal P}{}_{\cal C} \ ,
\eea
where we have defined ${\cal D}_{\cal A} = {\cal E}^{\cal M}{}_{\cal A}\, \partial_{\cal M}$, which are generalized diffeomorphism scalars and transform anomalously under extended Lorentz transformations
\be
\delta {\cal F}_{\cal A B C} = \xi^{\cal P} \partial_{\cal P} {\cal F}_{\cal A  B C} - 3
\left({\cal D}_{[{\cal A}} \Gamma_{{\cal B C}]} + \Gamma_{[{\cal A}}{}^{\cal D} \, {\cal F}_{{\cal B C}]{\cal D}}\right) \ .
\ee
As a consequence of the strong constraint \eqref{ExtendedStrongConstraint} and due to the linear and quadratic constraints for the gaugings \eqref{LinearQuadratic}, the generalized fluxes satisfy Bianchi identities
\be
\left[{\cal D}_{\cal A},\, {\cal D}_{\cal B}\right] = {\cal F}_{\cal A B}{}^{\cal C} \, {\cal D}_{\cal C} \ , \ \ \ \ {\cal D}_{[{\cal A}} {\cal F}_{{\cal B C D}]} - \frac 3 4 {\cal F}_{[{\cal A B}}{}^{\cal E} {\cal F}_{{\cal C D}]{\cal E}} = 0 \ . \label{extendedBI}
\ee
Generalized fluxes can also be defined in the double space
\bea
F_{A B C} = 3 D_{[A} E^M{}_B \, E^N{}_{C]} \, \eta_{M N} \ ,
\eea
where we defined double flat derivatives $D_A = E^M{}_A \partial_M$. They satisfy their own Bianchi identities
\be
\left[D_A,\, D_B\right] = F_{ A B}{}^{ C} \, {D}_{ C} \ , \ \ \ \ { D}_{[{ A}} { F}_{{ B C D}]} - \frac 3 4 { F}_{[{ A B}}{}^{ E} { F}_{{ C D}]{ E}} = 0 \ .
\ee
The extended fluxes can then be cast in terms of the double ones. The projections that are relevant to our discussion are
\begin{eqnarray}
{\cal F}_{\underline a\overline{b c} }&=& (\chi^{\frac12})_{\underline a}{}^{\underline e}\;
F_{\underline e\overline{b c} }\ ,\label{FabcGF}\\
{\cal F}_{\underline a \overline{b \gamma}}&=& - \left[(\chi^{\frac12})_{\underline a}{}^{\underline e}
\left(
{\cal E}_{\alpha}{}^{ \underline d} F_{\overline b\underline{d e}} + D_{\overline b} {\cal E}_{\alpha \underline e}\right)-D_{\overline b}(\Box^{\frac12})_{\alpha}{}^{\beta}\; {\cal E}_{\beta \underline a}\right] e^{\alpha}{}_{\bar \gamma}\ ,\label{FabgammaGF}\\
{\cal F}_{\underline a\overline{\alpha\beta} }&=&g\; f_{\delta\epsilon}{}^{\gamma} \; {\cal E}_{\gamma\,\underline a}\;(\Box^{\frac12})^{\delta}{}_{\alpha} (\Box^{\frac12})^{\epsilon}{}_{\beta}\;e^{\alpha}{}_{\overline{\alpha}} \,e^{\beta}{}_{\overline \beta}\label{FaalpbetGF} \\
&&+ (\chi^{\frac12})_{\underline a}{}^{\underline b}\; {\cal E}_{\alpha}{}^{\underline c}\;
 e^{\alpha}_{[\overline \alpha}\, e^{\beta}{}_{\overline \beta]}
 \left[ F_{\underline{bcd}} {\cal E}_{\beta}{}^{\underline d}
+ (2\, D_{\underline c} {\cal E}_{\beta\, \underline b}-D_{\underline b} {\cal E}_{\beta\, \underline c})\right]\nn \\
&&+  e^{\alpha}_{[\overline \alpha}\, e^{\beta}{}_{\overline \beta]}\;  D_{\underline b}(\Box^{\frac12})^{\gamma}{}_{\alpha} \left[
(\chi^{\frac12})_{\underline a}{}^{\underline b}(\Box^{\frac12})_{\beta}{}^{\gamma} +{\cal E}_{\gamma \underline a}\; {\cal E}_{\beta}{}^{\underline b}
 \right]\ .
\nn
\end{eqnarray}

\section{The generalized Bergshoeff-de Roo identification} \label{sec::gGG}

In the previous section we saw that the generalized frame in the extended space can undergo a $G$ and $H$ decomposition in terms of a double generalized frame $E_M{}^A$, fundamental $G$-vectors ${\cal A}_M{}^\alpha$ playing the role of gauge fields of the gauge group $\cal K$, and a $g$-valued matrix $e_\alpha{}^{\overline \alpha}$. Our intention here is to {\it lock} the extended dof ${\cal A}_M{}^\alpha$ and $e_\alpha{}^{\overline \alpha}$ in terms of (derivatives of) the double generalized frame $E_M{}^A$, leaving it as the unique dof of the theory.

To avoid detours, we include an appendix where we discuss different possibilities for locking the extended dof, also comparing with previous attempts. Here we go straight to the point. The identification we make is the following
\bea
{\cal K} = \overline {\cal H} \ .
\eea
A priori this would seem impossible because
\be
{\rm dim} {\cal K}  = k \ , \ \ \ \ \ \dim \overline {\cal H} = \frac {(D+k)(D+k-1)}{2} \ ,
\ee
and so there is no way to match both dimensions for finite $k$. However, the identification we make forces $k \to \infty$, so there is no conflict in making this choice as long as the dictionary between both groups is well established. Since the indices in ${\cal K}$ and $\overline {\cal H}$ are noted differently, we need to introduce a map between them
\be
 V_{\overline {\cal A}}{}^{\overline {\cal B}} = -g  \, V_\alpha \, (t^\alpha)_{\overline {\cal A}}{}^{\overline {\cal B}} \ , \label{IndexRelation}
\ee
where $\left(t_\alpha\right)_{\overline{\cal A}}{}^{\overline{\cal B}}$ denote the generators of the gauge algebra, and satisfy $\left[t_\alpha, t_\beta\right]=f_{\alpha\beta}{}^{\gamma} t_{\gamma}$.

The locking is accompanied by the gauge fixing discussed in the previous section, ${\cal A}_{\overline M \alpha} = 0$ and $e_\alpha{}^{\overline \alpha} = constant$, which required non vanishing $\Gamma_{\overline {a\alpha}}$ and $\Gamma_{\overline{\alpha \beta}}$ parameters \eqref{GaugedFixedParameters}. The only dof that one has to identify is then ${\cal E}_{\alpha \underline a} = E^M{}_{\underline a} {\cal A}_{M\alpha} \equiv {\cal A}_{\underline a \alpha}$,  that transforms as a projected generalized connection
\be
\delta {\cal A}_{\underline a \alpha} = \widehat {\cal L}_\xi {\cal A}_{\underline a \alpha} - {\cal D}_{\underline a} \xi_\alpha + g f_{\alpha \beta}{}^\gamma \xi^\beta {\cal A}_{\underline a \gamma} + {\cal A}_{\underline d \alpha} \Gamma^{\underline d}{}_{\underline {a}} \ . \label{TransformationA}
\ee
Using the map \eqref{IndexRelation}, this can be rewritten as
\begin{eqnarray}
 \delta {\cal A}_{\underline a \overline{\cal B C}}= \widehat {\cal L}_\xi {\cal A}_{\underline a \overline{\cal B C}} - {\cal D}_{\underline a} \xi_{\overline{\cal B C}} +  2 {\cal A}_{\underline a \overline{\cal D} [\overline{\cal C}}\, \xi^{\overline{\cal D}}{}_{\overline{\cal B}]} +  {\cal A}_{\underline d \overline{\cal B C}}\, \Gamma^{\underline{d}}{}_{\underline a} \ , \label{TransformationAH}
\end{eqnarray}
and this is precisely the way in which the following projection of the extended fluxes transforms
\bea
\delta {\cal F}_{\underline a \overline {\cal B C}} = \widehat {\cal L}_\xi {\cal F}_{\underline a \overline {\cal B C}} - {\cal D}_{\underline a} \Gamma_{\overline{\cal B C}} + 2 {\cal F}_{\underline a  \overline {\cal D} [\overline{\cal C}} \Gamma^{\overline{\cal D}}{}_{\overline{\cal B}]} + {\cal F}_{\underline d  \overline {\cal B C}} \Gamma^{\underline d}{}_{\underline {a}} \ . \label{TransformationFProjected}
\eea

The resemblance between \eqref{TransformationAH} and \eqref{TransformationFProjected} turns into an exact identity provided
\bea
\boxed{\begin{split}
\xi_{\overline {\cal A B}} &= -g  \, \xi_\alpha \, (t^\alpha)_{\overline {\cal A B}} \ =\ \Gamma_{\overline {\cal A B}}\ , \\
{\cal A}_{\underline a \overline {\cal B C}} &= -g  \, {\cal E}_{\alpha \underline a} \, (t^\alpha)_{\overline {\cal B C}} \ =\ {\cal F}_{\underline a \overline {\cal B C}}  \ . \end{split}}\label{Locking}
\eea
This is the key identity of the paper: it corresponds to an exact locking of the extended degrees of freedom, and it is what we call {\it the generalized Bergshoeff-de Roo identification}. Its supersymmetric extension will be discussed on a separate section.

We assume the generators to define invertible maps\footnote{This assumption might actually be too strong, as in the second identity the Kroneker delta must be replaced by the projector to the adjoint representation. We leave for future work to provide a more rigorous treatment of this infinite-dimensional mathematical structure.}
\begin{eqnarray}
\left(t^\alpha\right)_{\overline{\cal A} \overline{\cal B}} \left(t_\beta\right)^{\overline{\cal A} \overline{\cal B}}&=& X_R \,\delta^\alpha_{\beta}\ ,\label{GenPr1}\\
\left(t^\alpha\right)_{\overline{\cal A} \overline{\cal B}} \left(t_\alpha\right)^{\overline{\cal C} \overline{\cal D}}&=&X_R \;\delta_{\overline{\cal A} \overline{\cal B}}^{\overline{\cal C} \overline{\cal D}}
\ ,\label{GenPr2}
\end{eqnarray}
where $X_R$ denotes the Dynkin index of the representation. Notice that consistency of the equations above requires
\begin{eqnarray}
\kappa_{\alpha\beta}&=&-f_{\alpha\gamma}{}^{\delta}\, f_{\beta\delta}{}^{\gamma}=
 -\frac{4}{X_R}\, (t_{\gamma})^{\overline{\cal K}}{}_{[\overline{\cal C}}\,
  (t_{\alpha})_{\overline{\cal D}] \overline{\cal K}} (t_{\beta})^{\overline{\cal C L}}\,
  (t^{\gamma})_{\overline{\cal L} }{}^{\overline{\cal D} }= X_R \, (N-2)\, \kappa_{\alpha\beta}\ .
\end{eqnarray}
Hence $X_R=\frac{1}{N-2}$, where $N=\delta_{\overline{\cal A}}^{\overline{\cal A}}$. In the limit we consider this would vanish. Nevertheless it will play a fundamental role as a regulator of certain divergent traces and so, we will still keep track of this factor until the end of the computation.

It is worth noting that the identification (\ref{Locking}) holds independently of the gauge fixing conditions discussed in the previous section. Without such a gauge fixing the identification would generate infinitely many gauge dof ${\cal E}_{\alpha \overline a}$ and $e_\alpha{}^{\overline \alpha}$ with an infinite amount of gauge symmetry parameterized by $\Gamma_{\overline{\alpha a}}$ and $\Gamma_{\overline{\alpha \beta}}$. The gauge fixing eliminates the redundant gauge dof, and together with the locking leaves the double generalized frame $E_M{}^A$ as the unique dynamical field. In addition, \eqref{GaugedFixedParameters} defines $\Gamma_{\overline {a \alpha}}$  and $\Gamma_{\overline{\alpha\beta}}$ in terms of $\xi^\alpha$, which due to the identification depends again on $\Gamma_{\overline{\cal AB}}$. The iteration can be pursued order by order to obtain a derivative expansion. For later use we display here the first contributions to the $\alpha'\sim {\cal O}(g^{-2})$ expansion
\begin{eqnarray}
\Gamma_{\overline \alpha \overline a}\!\!&=&\!\!-\frac{1}{g\, X_R} e^{\beta}{}_{\overline \alpha}\, D_{\overline a}\Gamma_{\overline {\cal C}\overline {\cal D}} \, (t_{\beta})^{\overline {\cal C}\overline {\cal D}}
-\frac1 {2\, g^3\, X_R^2} \, e^{\beta}{}_{\overline \alpha} \, {\cal F}^{\underline c}{}_{\overline{\cal A}\overline{\cal B}} \, {\cal F}_{\underline c \overline{\cal C}\overline{\cal D}} \, D_{\overline a} \Gamma^{\overline {\cal C}\overline {\cal D}} \, (t_{\beta})^{\overline{\cal A}\overline{\cal B}} +{\cal O}\left(\frac{1}{g^5}\right)\ ,
\label{GammaAlphaAExpansion}\\
\Gamma_{\overline {\alpha \beta}}\!\!&=&\!\!-\frac{1}{X_R}\, f_{\alpha\beta}{}^{\gamma} \, \Gamma_{\overline{\cal A}\overline{\cal B}}\, (t_{\gamma})^{\overline{\cal A}\overline{\cal B}}\,  e^{\alpha}{}_{\overline \alpha}\,  e^{\beta}{}_{\overline \beta} - \frac{1}{g^2\, X_R^2}\, D^{\underline a} \Gamma_{\overline{\cal A}\overline{\cal B}}\, {\cal F}_{\underline a \overline{\cal C}\overline{\cal D}} \,
(t_{\alpha})^{\overline{\cal A}\overline{\cal B}} \, (t_{\beta})^{\overline{\cal C}\overline{\cal D}} \, e^{\alpha}{}_{[\overline \alpha} \, e^{\beta}{}_{\overline \beta]}
+{\cal O}\left(\frac{1}{g^4}\right)\ . \nn \\\label{GammaAlphaBetaExpansion}
\end{eqnarray}
Analogously we can solve iteratively for the relevant components of the generalized fluxes
\begin{eqnarray}
{\cal F}_{\underline a \overline{bc}} &=& F_{\underline a \overline{b c}} - \frac 1 {2\, X_R\, g^2} \, {\cal F}_{\underline a \overline {\cal CD}} \,{\cal F}^{\underline e \overline {\cal CD}}\, F_{\underline e \overline {bc}} + {\cal O}\left(\frac 1 {g^4}\right)\ , \\
{\cal F}_{\underline{a}\overline {b  \alpha}}&=&
\frac{1}{X_R\, g}\left( F_{\underline{a} \overline b \underline c} {\cal F}^{\underline{c} \overline{\cal A} \overline{\cal B}}+ D_{\overline b} {\cal F}_{\underline{a}}{}^{\overline{\cal A} \overline{\cal B}} \right)
e_\alpha{}^{\overline \alpha} (t^\alpha)_{\overline{\cal A} \overline{\cal B}}+\;{\cal O}\left(\frac{1}{g^3}\right) \ ,
\label{Fabalpha}\\
{\cal F}_{\underline{a} \overline {\alpha \beta}}&=&
-\frac{1}{X_R} \, f_{\alpha \beta\gamma}e^{\alpha}{}_{\overline{\alpha}} \, e^{\beta}{}_{\overline \beta} \, {\cal F}_{\underline{a} \overline{\cal C} \overline{\cal D}}\, (t^\gamma)^{\overline{\cal C} \overline{\cal D}}\nn\\
&& + \frac{1}{X_R^2\, g^2} \,\left( F_{\underline{acd}}\, {\cal F}^{\underline{c}}{}_{\overline{\cal A} \overline{\cal B}} \,
{\cal F}^{\underline{d}}{}_{\overline{\cal C} \overline{\cal D}}
+\left(D_{\underline a} {\cal F}_{\underline{c} \overline{\cal A} \overline{\cal B}}
-2\, D_{\underline c} {\cal F}_{\underline{a} \overline{\cal A} \overline{\cal B}} \right)
{\cal F}^{\underline{c}}{}_{\overline{\cal C} \overline{\cal D}}\right)
e^{\alpha}{}_{[\overline{\alpha}} \, e^{\beta}{}_{\overline \beta]} (t_\alpha)^{\overline{\cal A} \overline{\cal B}}
(t_\beta)^{\overline{\cal C} \overline{\cal D}}\nn \\
&& + \frac{1}{X_R^3\, g^2}\,  f_{\beta\gamma\delta} \, {\cal F}_{\underline a \overline{\cal A} \overline{\cal B}} \, {\cal F}_{\underline c\; \overline{\cal C} \overline{\cal D}} \, {\cal F}^{\underline c}{}_{\overline{\cal E} \overline{\cal F}} \,
e^{\alpha}{}_{[\overline{\alpha}} \, e^{\beta}{}_{\overline \beta]}\,
(t^\gamma)^{\overline{\cal A} \overline{\cal B}}
(t^\delta)^{\overline{\cal C} \overline{\cal D}} (t_\alpha)^{\overline{\cal E} \overline{\cal F}}\;\;
+\, {\cal O}\left(\frac{1}{g^4}\right)\ .
\label{Faalphabeta}
\end{eqnarray}

Let us conclude with some words on how the extended space becomes infinite after the identification. First we identify $\cal K$ indices $\alpha$ with $\overline {\cal H}$ indices $\overline {\cal A B}$, through the generators $(t_\alpha)_{\overline {\cal A B}}$. The $\overline {\cal H}$ indices now split under an $\overline H$ decomposition as $\overline {\cal A} = (\overline a,\, \overline \alpha)$, such that $\alpha \to \overline {\cal A B} \to (\overline {a b},\, \overline {a \beta},\, \overline {\alpha b},\, \overline {\alpha \beta})$. We have also introduced a bijective map $e: \overline h \to g$ in \eqref{bijection} that allows to convert $\overline h$ indices $\overline \alpha$ back to ${\cal K}$ indices $\alpha$, such that $\alpha \to \overline {\cal A B} \to (\overline {a b},\, \overline {a \beta},\, \overline {\alpha b},\, \overline {\alpha \beta}) \to (\overline {a b},\, \overline {a} \beta,\,\alpha \overline {b},\, \alpha \beta) $. Repeating this procedure over and over leads to an infinite dimensional tangent space. As we will see, interestingly, contractions in this space converge yielding finite order by order expressions. We show the first and second order expansion in what follows.

\subsection{First order}

We show here that the exact identification (\ref{Locking}), when expanded to first order in $\alpha' \sim {\cal O}\left(g^{-2}\right)$, reproduces the expected first-order generalized Green-Schwarz transformation of \cite{Marques:2015vua}. To this end, we begin with the $\overline H$ projection of the the generalized frame, whose exact transformation was written in \eqref{TransformationEoverlineA}
\be
\delta E_M{}^{\overline a} = \widehat {\cal L}_\xi E_M{}^{\overline a} + E_M{}^{\overline b}\, \Lambda_{\overline b}{}^{\overline a} + D^{\overline a} \xi^\alpha {\cal C}_{M \alpha} \ .
\ee
In \eqref{RedefinitionVectors} the gauge vector ${\cal C}_M{}^\alpha$ was related to another one ${\cal A}_M{}^\alpha$ and these dof were locked in \eqref{Locking} by identifying them with  the generalized flux component ${\cal F}_{\underline a \overline {\cal B C}}$. Implementing these relations  yields
\be
\delta E_M{}^{\overline a} = \widehat {\cal L}_\xi E_M{}^{\overline a} + E_M{}^{\overline b}\, \Lambda_{\overline b}{}^{\overline a}
														- \frac{1}{g^2 X_R} E_{M}{}^{\underline b} \;{\cal F}_{\underline b \overline{\cal C}\overline{\cal D}} \;D^{\overline a} \Gamma^{\overline{\cal C}\overline{\cal D}}  +{\cal O}\left(\frac{1}{g^4}\right) \ . \label{deltaEoverline}
\ee
The contracted factor on the last term splits as follows
\begin{eqnarray}
{\cal F}_{\underline b \overline{\cal C}\overline{\cal D}}\; D^{\overline a} \Gamma^{\overline{\cal C}\overline{\cal D}} =
{\cal F}_{\underline b \overline {c  d}} \;D^{\overline a} \Gamma^{\overline {c d}} + {\cal F}_{\underline b \overline {\alpha \beta }} \;D^{\overline a} \Gamma^{\overline {\alpha  \beta}}
+2 {\cal F}_{\underline b \overline {\alpha c}} \;D^{\overline a} \Gamma^{\overline {\alpha c}} \ .
\label{FDG}
\end{eqnarray}
In the last term the mixed contraction $2\,{\cal F}_{\underline b \overline {\alpha c}} \; D^{\overline a} \Gamma^{\overline {\alpha c}}\sim {\cal O}\left(g^{-2}\right)$ by virtue of \eqref{GammaAlphaAExpansion}, \eqref{Fabalpha} and so it can be ignored at this stage. For the second term we need to take the leading order from \eqref{Faalphabeta} and \eqref{GammaAlphaBetaExpansion}, obtaining
\begin{eqnarray}
{\cal F}_{\underline b \overline {\alpha \beta} }\; D^{\overline a} \Gamma^{\overline {\alpha  \beta}}=  \frac{1}{X_R} {\cal F}_{\underline{b} \overline{\cal C} \overline{\cal D}}\ D^{\overline{a}} \Gamma^{\overline{\cal C}\overline{\cal D}} +{\cal O}\left(\frac{1}{g^2}\right)  \ .
\label{FDGalphabeta}
\end{eqnarray}
Remarkably the RHS of (\ref{FDGalphabeta}) contains the exact same expression appearing in the LHS of \eqref{FDG}. Then, replacing \eqref{FDGalphabeta} in \eqref{FDG} we read off
\be
{\cal F}_{\underline b \overline{\cal C}\overline{\cal D}}\; D^{\overline a} \Gamma^{\overline{\cal C}\overline{\cal D}}
= \frac{X_R}{-1+X_R}\, {\cal F}_{\underline b \overline {cd}} \;D^{\overline a} \Gamma^{\overline {c d}} +{\cal O}\left(\frac{1}{g^2}\right)
= \frac{X_R}{-1+X_R}\,  F_{\underline b \overline {c d}} \;D^{\overline a} \Gamma^{\overline {c d}} +{\cal O}\left(\frac{1}{g^2}\right) \ ,\label{trick}
\ee
where in the last equality we used the fact that extended and double fluxes are equal to leading order (\ref{FabcGF}). Then (\ref{deltaEoverline}) becomes
\begin{eqnarray}\boxed{
\delta E_{M}{}^{\overline a}=  \widehat {\cal L}_\xi E_M{}^{\overline a} + E_M{}^{\overline b}\, \Lambda_{\overline b}{}^{\overline a} - \frac b 2  \, E_M{}^{\underline d} F_{\underline d \overline {b c}} \, D^{\overline a}  \Lambda^{\overline {bc}} +{\cal O}\left(\frac{1}{g^4}\right)} \ , \label{deltaEoverline1}
\end{eqnarray}
were we defined
\begin{eqnarray}
b = \frac{2}{g^2 (-1+X_R)}\ ,\label{abmatch}
\end{eqnarray}
and used \eqref{LorentzParamterRedefinition} to rename the Lorentz parameter. We then find that the first non-trivial order in $\alpha'$ \eqref{deltaEoverline1} exactly reproduces the first order generalized Green-Schwarz transformation of DFT  \cite{Marques:2015vua}.

Let us now focus on the transformation of the $\underline H$ projection (\ref{TransformationEunderlineA})
\be\delta E_M{}^{\underline a} = \widehat {\cal L}_\xi E_M{}^{\underline a} + E_M{}^{\underline b}\, \Lambda_{\underline b}{}^{\underline a} - \partial_{\overline M} \xi^\alpha \, {\cal C}_{Q \alpha} \, E^{Q \underline a} \ . \label{TransfUnderlineH}
\ee
To first order in $\alpha'$ we find
\be
 - \partial_{\overline M} \xi^\alpha \, {\cal C}_{Q \alpha} \, E^{Q \underline a} = \frac 1 {g^2 X_R} \partial_{\overline M} \Gamma^{\overline{\cal C D}} {\cal F}^{\underline a}{}_{\overline{\cal C D}} + {\cal O}\left(\frac 1 {g^4}\right)\ , \label{CtoFlux}
\ee
and in addition we can decompose
\be
\partial_{\overline M} \Gamma^{\overline{\cal C D}} {\cal F}^{\underline a}{}_{\overline{\cal C D}} = \partial_{\overline M} \Gamma^{\overline{cd}} {\cal F}^{\underline a}{}_{\overline{cd}} + \partial_{\overline M} \Gamma^{\overline{\alpha \beta}} {\cal F}^{\underline a}{}_{\overline{\alpha \beta}} + 2 \partial_{\overline M} \Gamma^{\overline{\alpha c}} {\cal F}^{\underline a}{}_{\overline{\alpha c}} \ . \label{DecompUnderlineH}
\ee
Again the last term can be neglected as it contributes to higher order, and the second one gives
\be
 \partial_{\overline M} \Gamma^{\overline{\alpha \beta}} {\cal F}^{\underline a}{}_{\overline{\alpha \beta}} = \frac 1 {X_R} \partial_{\overline M} \Gamma^{\overline{\cal C D}} {\cal F}^{\underline a}{}_{\overline{\cal C D}} + {\cal O}\left(\frac 1 {g^2}\right) \ ,
\ee
which reinserting into \eqref{DecompUnderlineH} and replacing by double fluxes yields
\be
\partial_{\overline M} \Gamma^{\overline{\cal C D}} {\cal F}^{\underline a}{}_{\overline{\cal C D}} = \frac{X_R}{1-X_R} \partial_{\overline M} \Gamma^{\overline{c d}} F^{\underline a}{}_{\overline{cd}}  + {\cal O}\left(\frac 1 {g^2}\right) \ , \label{trick2}
\ee
in analogy with \eqref{trick}. We can now replace \eqref{trick2} into \eqref{CtoFlux} into \eqref{TransfUnderlineH}, and use the leading order redefinition of Lorentz parameters \eqref{LorentzParamterRedefinition} to arrive at
\be\boxed{\delta E_M{}^{\underline a} = \widehat {\cal L}_\xi E_M{}^{\underline a} + E_M{}^{\underline b}\, \Lambda_{\underline b}{}^{\underline a} + \frac b 2 \partial_{\overline M} \Lambda^{\overline {b c}} \, F^{\underline a}{}_{\overline {b c}}  + {\cal O}\left(\frac 1 {g^4}\right)} \ . \label{deltaEunderline1}
\ee
This matches the other projection of the first order generalized Green-Schwarz transformation \cite{Marques:2015vua}.

As previously anticipated, the final result is finite and non vanishing in the limit $X_R\to 0$ leading to the simple identification $b\to-2/g^2$.

\subsection{Second order}

Now we obtain the so far unknown ${\cal O}(\alpha'{}^2)$ deformations of the generalized Green-Schwarz transformation. Once again we start considering the $\overline H$ projection of the generalized frame \eqref{TransformationEoverlineA}
\bea
\delta E_M{}^{\overline a} &=& \widehat {\cal L}_\xi E_M{}^{\overline a}\, + \,E_M{}^{\overline b}\, \Lambda_{\overline b}{}^{\overline a} - \frac{1}{g^2 X_R} E_{M}{}^{\underline c} \;(\chi^{-\frac12})_{\underline c}{}^{\underline b} \;{\cal F}_{\underline b \overline{\cal C}\overline{\cal D}} \;D^{\overline a} \Gamma^{\overline{\cal C}\overline{\cal D}}\;.  \label{deltaEoverline2}
\eea
After repeating the steps of the previous section for the last term but keeping the following order terms that were previously neglected, gives
\begin{eqnarray}
{\cal F}_{\underline b \overline{\cal C}\overline{\cal D}} \;D^{\overline a} \Gamma^{\overline{\cal C}\overline{\cal D}}&=&
\frac{X_R}{(-1+X_R)}\;{\cal F}_{\underline b \overline{c d}} \;D^{\overline a} \Gamma^{\overline{c d}} \label{FDGamma} \\ &&-\;\frac{2}{g^2\,(-1+X_R)}  \left[ \;{\cal F}_{\underline b}{}_{\overline{\cal E F}} {\cal F}_{\underline c}{}_{\overline{\cal G}}{}^{\overline{\cal F}}\left(
{\cal F}^{\underline c \overline{\cal C G}} D^{\overline a} \Gamma_{\overline{\cal C}}{}^{\overline{\cal E} }
 -  {\cal F}^{\underline c \overline{\cal C E}} D^{\overline a}\Gamma_{\overline{\cal C}}{}^{\overline{\cal G} }\right)
\right.\nn\\
&&\;\;\;\;-\;\left(D^{\overline a}\Gamma^{\overline{\cal EG}}\right) \left(F_{\underline{ b  c d}}{\cal F}^{\underline c}{}_{\overline{\cal E F}} {\cal F}^{\underline d}{}_{\overline{\cal G}}{}^{\overline{\cal F}}
+ D_{\underline b}{\cal F}^{\underline c}{}_{\overline{\cal  EF}} {\cal F}_{\underline c}{}_{\overline{\cal G}}{}^{\overline{\cal F}}  - 2\; D_{\underline c}  {\cal F}_{\underline b}{}_{\overline{\cal E F}}
{\cal F}^{\underline c}{}_{\overline{\cal G}}{}^{\overline{\cal F}}   \right)\nn\\
 && \;\;\;\; -\left.\,{\cal F}_{\underline  b\overline{\cal E F} }D^{\overline a} \left(D^{\underline c}\Gamma^{\overline{\cal E G}}{\cal F}_{\underline c \overline{\cal G}}{}^{\overline{\cal F}} \right)-\; D^{\overline a} D^{\overline c}  \Gamma^{\overline{\cal E F}}\left(F_{\overline c\underline{d b}} {\cal F}^{\underline d}{}_{\overline{\cal  EF}} + D_{\overline{c}} {\cal F}_{\underline b}{}_{\overline{\cal  EF}}         \right) \right] \nn \\ && + {\cal O}\left(g^{-6}\right) \ . \nn
\end{eqnarray}
Notice that the $\chi$ factor in (\ref{deltaEoverline2}) has the effect of switching the extended into the double fluxes $(\chi^{-\frac12})_{\underline a}{}^{\underline b}\;{\cal F}_{\underline b \overline{c d}}= \;F_{\underline a \overline{c d}}$ in the first line of (\ref{FDGamma}) whereas it is equivalent to a Kronecker delta on the second to fourth lines,  at this order.

We can try to reproduce the logic of the first order computation, namely to split indices ${\overline {\cal A}}\to ({\overline a, \overline \alpha})$, replace (\ref{GammaAlphaAExpansion})-(\ref{Faalphabeta}), discard terms with mixed indices (as they are subleading) and get rid of terms with Greek dummy indices by noting that they are proportional to the same terms with $\overline {\cal H}$ indices. This procedure turns out not to work for each individual term at ${\cal O}\left(g^{-4}\right)$ but remarkably it does for the whole sum in (\ref{FDGamma}). Hence, we conclude

\bea\boxed{\begin{split}
\delta E_{M}{}^{\overline a}= & \ \widehat {\cal L}_\xi E_M{}^{\overline a} + E_M{}^{\overline b}\, \Lambda_{\overline b}{}^{\overline a} - \frac b 2  \, E_M{}^{\underline d} F_{\underline d \overline {b c}} \, D^{\overline a}  \Lambda^{\overline {bc}}  \\
&-\, \frac12\; b^2\; E_M{}^{\underline b}\left[
														D^{\overline{a}}D^{\overline c}\Lambda^{\overline{e f}}\left(F_{\overline c\underline{d b}} F^{\underline d}{}_{\overline{e f}} + D_{\overline c} F_{\underline b \overline{e f}}\right)	
														\right.
-F_{\underline b \overline{e f}} F_{\underline c \overline d}{}^{\overline f}\left(
F^{\underline c}{}^{\overline{h d}} D^{\overline a}\Lambda_{\overline h}{}^{\overline e}
-F^{\underline c}{}^{\overline{h e}} D^{\overline a}\Lambda_{\overline h}{}^{\overline d}
\right) \\
&\ \ \ \ \ \ \ \ \ \ \ \ \ \ \  +\;\left. F^{\underline c}{}_{\overline{e f}}\;D^{\overline a}\Lambda^{\overline{e}}{}_{\overline g} \left(F_{\underline {b c d}} F^{\underline d \overline{g f}}
-D_{\underline b} F_{\underline c}{}^{\overline{g f}} + 2\; D_{\underline c} F_{\underline b}{}^{ \overline{g f}}
\right)
+F_{\underline b \overline{e f}}D^{\overline a}\left(D^{\underline c}\Lambda^{\overline{e d}} F_{\underline c \overline{d}}{}^{\overline f}\right)						
							\right]
\\ & +\, {\cal O}\left(g^{-6}\right)\ .\end{split}} \nn
\eea
\be
\label{Second1}
\ee
It is quite remarkable that all the dependence on the Dynkin coefficient $X_R$ and the coupling constant $g$ has arranged once again in the same parameter $b$ as before \eqref{abmatch}.

The transformation of the $\underline H$ projection (\ref{TransformationEunderlineA}) is completely analogous
\bea
\delta E_M{}^{\underline a} &=& \widehat {\cal L}_\xi E_M{}^{\underline a} + E_M{}^{\underline b}\, \Lambda_{\underline b}{}^{\underline a} + \frac{1}{g^2 \, X_R}\partial_{\overline M}
														 \Gamma^{\overline{\cal C}\overline{\cal D}}
														\;(\chi^{-\frac12})^{\underline a}{}_{\underline b}\;{\cal F}^{\underline b}{}_{\overline{\cal C}\overline{\cal D}}\ .
\end{eqnarray}
Repeating the procedure for the previous projection one readily arrives at
\bea \boxed{\begin{split}
\delta E_M{}^{\underline a} =& \ \widehat {\cal L}_\xi E_M{}^{\underline a} + E_M{}^{\underline b}\, \Lambda_{\underline b}{}^{\underline a} + \frac b 2 \partial_{\overline M} \Lambda^{\overline {b c}} \, F^{\underline a}{}_{\overline {b c}} \\
& +\frac12\; b^2\; E_M{}^{\overline b}\left[
D_{\overline{b}}D^{\overline c}\Lambda^{\overline{e f}}\left(F_{\overline c\underline d}{}^{\underline a} F^{\underline d}{}_{\overline{e f}} +
D_{\overline c} F^{\underline a}{}_{\overline{e f}}\right)	\right.
-F^{\underline a}{}_{\overline{e f}} F_{\underline c \overline d}{}^{\overline f}\left(
F^{\underline c}{}^{\overline{h d}} D_{\overline b}\Lambda_{\overline h}{}^{\overline e}
-F^{\underline c}{}^{\overline{h e}} D_{\overline b}\Lambda_{\overline h}{}^{\overline d}
\right) \\
&\ \ \ \ \ \ \ \ \ \ \ \ \ + \left. F^{\underline c}{}_{\overline{e f}}\;D_{\overline b}\Lambda^{\overline{e}}{}_{\overline g} \left(F^{\underline a}{}_{\underline{c d}} F^{\underline d \overline{g f}}
-D^{\underline a} F_{\underline c}{}^{\overline{g f}} + 2\; D_{\underline c} F^{\underline a  \overline{g f}}\right)
+F^{\underline a}{}_{\overline{e f}}D_{\overline b}\left(D^{\underline c}\Lambda^{\overline{e d}} F_{\underline c \overline{d}}{}^{\overline f}\right)\right]	\\
& +\;{\cal O}\left(g^{-6}\right)  \ . \end{split}}\nn
\eea
\be
\label{Second2}
\ee

As a non-trivial check we have verified closure. The brackets receive second order corrections, and are given by\footnote{\label{typo}We thank F. Hassler and A. Gitsis for identifying a typo in Equation (\ref{bracket}).}
\bea
\xi_{12}^M&=& 2 \xi_{[1}^{P} \partial_{P} \xi_{2]}^{M} + \partial^{M} \xi_{[1}^{P} \xi_{2]{P}}  \nn\\&&
- \frac{b}{2}\; \Lambda_{[1}^{\overline{cd}} \; \partial^M \Lambda_{2]}{}_{\overline{cd}}
\;+\;b^2\left[ \partial^M\Lambda_{[1}^{\overline{ef}}\; D^{\underline c}\Lambda_{2]\overline e}{}^{\overline d} F_{\underline c\overline{df}}\;+\;\frac12 \partial^M\left(D^{\overline c}\Lambda_{[1}^{\overline{ef}}\right)D_{\overline{c}} \Lambda_{2]}{}_{\overline{ef}}   \right]+{\cal O}(b^3)\ , \nn \\
\Lambda_{12}^{\overline {ab}}&=& 2\;\xi_{[1}^N\partial_N\Lambda_{2]}^{\overline{ab}}\,-\,2\;\Lambda_{[1}^{\overline{ac}}\,\Lambda_{2]}{}_{\overline c}{}^{\overline b}
+ \,b\; D^{\overline a}\Lambda_{[1}^{\overline{cd}}  \;D^{\overline b}\Lambda_{2]}{}_{\overline{cd}}\;+\;  b^2 \left[ \frac12 D^{\overline a}\Lambda_{[1}^{\overline{cd}}\;  D^{\overline b}\Lambda_{2]}^{\overline{ef}} F^{\underline g}{}_{\overline{cd}}\; F_{\underline{g}\overline{ef}}\right. \nn \\
&& +\left. D^{\overline a} D^{\overline e}\Lambda_{[1}^{\overline{cd}}\;  D^{\overline b} D_{\overline e}\Lambda_{2]}{}_{\overline{cd}}
+\,2\;  D^{[\overline a}\left(D^{\underline c}\Lambda_{[1}^{\overline{ed}}\; F_{\underline c\overline d}{}^{\overline f} \right)\;  D^{\overline b]}\Lambda_{2]}{}_{\overline{ef}} \right]+{\cal O}(b^3)\ , \label{bracket} \\
\Lambda_{12}^{\underline {ab}}&=& 2\;\xi_{[1}^N\partial_N\Lambda_{2]}^{\underline{ab}}\,-\,2\;\Lambda_{[1}^{\underline{ac}}\,\Lambda_{2]}{}_{\underline c}{}^{\underline b}
+ \,b\; D^{\underline a}\Lambda_{[1}^{\overline{cd}}  \;D^{\underline b}\Lambda_{2]}{}_{\overline{cd}}\;+\;  b^2 \left[ \frac12 D^{\overline c}\Lambda_{[1}^{\overline{ef}}\;  D_{\overline c}\Lambda_{2]}^{\overline{gh}} F^{\underline a}{}_{\overline{ef}}\; F^{\underline{b}}{}_{\overline{gh}} \right. \nn \\
&&+ \left. D^{\underline a} D^{\overline e}\Lambda_{[1}^{\overline{cd}}\;  D^{\underline b} D_{\overline e}\Lambda_{2]}{}_{\overline{cd}}
+\,2\;  D^{[\underline a}\left(D^{\underline c}\Lambda_{[1}^{\overline{ed}}\; F_{\underline c\overline d}{}^{\overline f} \right)\;  D^{\underline b]}\Lambda_{2]}{}_{\overline{ef}} \right]+{\cal O}(b^3)\ . \nn
\eea
The first order exactly reproduces \cite{Marques:2015vua} and the second order is a new result. These expressions could also be obtained from the extended brackets \eqref{extendedBrackets} after performing the identification, but one has to take care of the fact that the identification introduces field dependence in the parameter components, which must be properly accounted for.

\section{Supersymmetry}\label{sec::susy}

We now consider the ${\cal N} = 1$ and $D=10$ supersymmetric formulation of extended gauged DFT \cite{Hohm:2011nu},\cite{Jeon:2011sq}. The fermionic degrees of freedom are two Majorana spinors: an extended generalized gravitino $\Psi_{\overline{\cal{A}}}$ (which is an $\overline {\cal H}$ vector and an $\underline {\cal H}$ spinor) that contains the gravitino $\Psi_{\overline a}$ and gauginos  $\Psi_{\overline \alpha}$ from the point of view of the double space, and a generalized dilatino $\rho$ (which is an $\overline {\cal H}$ singlet and an $\underline {\cal H}$ spinor) which will be ignored as it plays no relevant role in our analysis. Both are scalars under extended generalized diffeomorphisms and $\cal G$ invariant. The supersymmetry parameter $\epsilon$ is also a Majorana $\underline {\cal H}$ spinor. The gamma matrices satisfy a Clifford algebra for $\underline {\cal H}$
\be
\left\{ \gamma^{\underline{a}},\gamma^{\underline{b}} \right\} = 2 P^{\underline{a} \underline{b}}\ , \label{Cliff}
\ee
and we use the standard convention for antisymmetrization of $\gamma$-matrices $\gamma^{\underline{a \dots b}}=\gamma^{[\underline{a}} \dots \gamma^{\underline{b}]}$.  The Clifford relation \eqref{Cliff} implies the following useful identities
\begin{align}\label{iden}
 \begin{split}
\gamma_{\underline a} \gamma_{\underline b} &= \gamma_{\underline {a b}} + P_{\underline {ab}} \ , \\
\gamma_{\underline {a b}} \gamma_{\underline c} &=  \gamma_{\underline {a b c}} + 2 \gamma_{[\underline a} P_{\underline b] \underline c} =\gamma_{\underline a}\gamma_{\underline {bc}}+2\gamma_{[\underline a}P_{\underline c]\underline b}\ , \\
  \gamma_{\underline {a b}} \gamma^{\underline {c d}} & =  \gamma_{\underline {a b}}{}^{\underline {c d}} + 4 \gamma_{[\underline a}{}^{[\underline d} P_{\underline b]}{}^{\underline c]}
+ 2 P_{[\underline b}{}^{[\underline c}\, P_{\underline a]}{}^{\underline d]} \ .\\
 \end{split}
\end{align}

Crucial to the analysis is the derivative
\be
\nabla_{\cal A} V_{\cal B}= \mathcal{D}_{\cal A} V_{\cal B} - \omega_{\cal A B }{}^{\cal C} V_{\cal C}\ ,
\ee
which is $\cal H$ covariant provided the generalized spin connection transforms as follows
\be
\delta_{\Gamma} \omega_{\cal A B C} = - {\cal D}_{A} \Gamma_{\cal B C} + \omega_{\cal DBC} \Gamma^{\cal D}{}_{\cal A} + \omega_{\cal ADC} \Gamma^{\cal D}{}_{\cal B} + \omega_{\cal ABD} \Gamma^{\cal D}{}_{\cal B} \ .
\label{spinconnecttransf}
\ee
Compatibility with the $\cal H$ invariants and vanishing generalized torsion impose constraints on the connection
\be
\omega_{{\cal A}({\cal B C})} = 0 \ , \ \ \ \omega_{{\cal A} \underline b \overline {\cal C}} =  0 \ ,
\ee
and generalized frame compatibility determines some projections of the connection in terms of the dynamical fields
\be
3 \, \omega_{[{\cal A B C}]} = {\cal F}_{\cal ABC}\ .
\ee
Together these imply
\be
\omega_{\underline a  \overline{\cal B C}} = {\cal F}_{\underline a  \overline{\cal B C}}\ , \ \ \ \omega_{\overline{\cal A}  \underline{b c}} = {\cal F}_{\overline{\cal A}  \underline{b c}} \ .
\ee
There are additional relations involving the generalized dilaton, but we ignore them here as they have no relevance in the analysis.

Let us now move on to the covariant derivative of a spinorial object. When we consider spinors, the covariant derivative takes an extra contribution. For example, the covariant derivative of the gravitino and the adjoint gravitino are
\be
\nabla_{\cal A} \Psi_{\overline{\cal B}}={\cal D}_{\cal A} \Psi_{\overline{\cal B}} - \omega_{{\cal A} \overline{\cal B}}{}^{\overline{\cal C}} \Psi_{\overline{\cal C}} - \frac{1}{4} \omega_{{\cal  A} \underline{bc}} \gamma^{\underline{bc}} \Psi_{\overline{\cal B}} \ , \quad \nabla_{\cal A} \overline{\Psi}_{\overline{\cal B}}={\cal D}_{\cal A} \overline{\Psi}_{\overline{\cal B}} - \omega_{{\cal A} \overline{\cal B}}{}^{\overline{\cal C}} \overline{\Psi}_{\overline{\cal C}} + \frac{1}{4} \omega_{{\cal  A} \underline{bc}} \, \overline{\Psi}_{\overline{\cal B}}\gamma^{\underline{bc}}  \ , \nn
\ee
where the adjoint spinor is defined through $\bar{\Psi}=\Psi^{t} C$ and the charge conjugation matrix satisfies
\bea
 C^{-1}=C^t=-C \ , \quad C \gamma C^{-1} = - \gamma^{t}\ .
\eea

We will only work to leading order in fermions, such that supersymmetric transformations of bosons are at most quadratic in fermions, and supersymmetric transformations of fermions are linear in fermions. On top of the extended generalized diffeomorphisms and $\cal H$ transformations, the extended generalized frame receives supersymmetric transformations given by
\bea
\delta_\epsilon \mathcal{E}_{\mathcal{M}}{}^{\underline a}&=&  -\frac 1 2 \,\bar{\epsilon}\, \gamma^{\underline a}\, \Psi_{\overline {\cal B}} \, \mathcal{E}_{\mathcal{M}}{}^{\overline {\cal B} }\ ,  \nn\\
\delta_\epsilon \mathcal{E}_{\mathcal{M}}{}^{\overline{a}} &=& \frac 1 2 \, \bar{\epsilon} \, \gamma^{\underline b}\,  \Psi^{\overline a} \, \mathcal{E}_{\mathcal{M} \underline b}\ , \label{susytransf} \\
\delta_\epsilon \mathcal{E}_{\mathcal{M}}{}^{\overline{\alpha}} &=& \frac 1 2 \, \bar{\epsilon} \, \gamma^{\underline b}\,  \Psi^{\overline \alpha} \, \mathcal{E}_{\mathcal{M} \underline b}\ .  \nn
\eea
The gravitino and gaugino on the other hand transform as follows
\bea
  \delta \Psi_{\overline a} &=& \xi^{\mathcal{M}} \partial_{\mathcal{M}} \Psi_{\overline a} + \Psi_{\overline {\cal B}} \Gamma^{\overline {\cal B}}{}_{\overline a} - \frac{1}{4} \Gamma_{{\underline{b} \underline{c}}} \gamma^{\underline{b} \underline{c}} \Psi_{\overline a} + \nabla_{\overline a} \epsilon  \ ,\\
    \delta \Psi_{\overline \alpha} &=& \xi^{\mathcal{M}} \partial_{\mathcal{M}} \Psi_{\overline \alpha} + \Psi_{\overline {\cal B}} \Gamma^{\overline {\cal B}}{}_{\overline \alpha} - \frac{1}{4} \Gamma_{{\underline{b} \underline{c}}} \gamma^{\underline{b} \underline{c}} \Psi_{\overline \alpha} + \nabla_{\overline \alpha} \epsilon\ .
\eea
The composition of these transformations closes to leading order in fermions with respect to the following brackets
\bea
\xi^{\cal M}_{12} &=& 2 \xi_{[1}^{\cal P} \partial_{\cal P} \xi_{2]}^{\cal M} + \partial^{\cal M} \xi_{[1}^{\cal P} \xi_{2]{\cal P}} + g \, f_{\cal N P}{}^{\cal M} \xi_1^{\cal N}  \xi_2^{\cal P}  - \frac{1}{2} {\cal E}^{\cal M}{}_{\underline a} \, \bar{\epsilon}_1\, \gamma^{\underline a}\, \epsilon_{2}\ , \nn\\
\Gamma_{12 {\cal A B}} &=& 2 \xi_{[1}^{\cal P} \partial_{\cal P} \Gamma_{2]{\cal A B}} + \Gamma_{[1 {\cal A}}{}^{\cal C} \, \Gamma_{2] {\cal B C}}  \ , \label{par0} \\
\epsilon_{12} &=& 2 \xi_{[1}^{\cal P} \partial_{\cal P} \epsilon_{2]} - \frac{1}{2} \Gamma_{[1 \underline {ab}} \gamma^{\underline{ab}} \epsilon_{2]} \ .\nn
\eea

We now proceed as before making the same $G \in {\cal G}$ and $H\in {\cal H}$ decomposition and gauge fixing. Imposing $\mathcal{E}_{\alpha}{}^{\bar{a}}=0$ and $\delta e_{\alpha}{}^{\bar{\alpha}}=0$ now gives a supersymmetric completion of the locked $\overline {\cal H}$ gauge parameters \eqref{GammaAlphaAExpansion}, \eqref{GammaAlphaBetaExpansion}
\bea
\Gamma_{\overline {\alpha a}}\!\! &=& \!\!e^\alpha{}_{\overline \alpha} \, (\Box^{- \frac 1 2}){}_\alpha{}^\beta\, \left(\partial_P \xi_\beta \, E^P{}_{\bar a} - \frac12 \, \bar{\epsilon}\, \gamma^{\underline{b}}\, \Psi_{\overline{a}} \, E^M{}_{\underline b}\, {\cal A}_{M \alpha} \right)\ ,\\
\Gamma_{\overline {\alpha \beta}}\!\! &=&\!\!  e^\alpha{}_{[\overline \alpha} \, e^\beta{}_{\overline \beta]} \, (\Box^{-\frac 1 2}){}_\alpha{}^\gamma \left(\delta (\Box^{\frac 1 2}){}_{\gamma \beta} - \partial^P \xi_\gamma\, {\cal A}_{P \beta} - g\, f_{\gamma \delta}{}^\lambda \, \xi^\delta \,(\Box^{\frac 1 2}){}_{\lambda \beta} - \frac 1 2 \, \bar\epsilon  \, \gamma^{\underline b}\, \Psi_{\underline \delta} \, e_\beta{}^{\underline \delta} \, E^M{}_{\underline b} {\cal A}_{M \gamma}\right)
\ . \nn
\eea

We now want to lock the extended dof (${\cal A}_{\underline a \alpha}$ and $\Psi_{\overline \alpha}$) in terms of the double fields $E_M{}^A$ and $\Psi_{\overline a}$. We have a starting point based on the pure bosonic locking, so lets begin by exploring wether that identification survives supersymmetry or not. On the one hand, the $G$-vectors transform as
\be
\delta {\cal A}_{\underline a \alpha} = \widehat {\cal L}_\xi {\cal A}_{\underline a \alpha} - {\cal D}_{\underline a} \xi_\alpha + g f_{\alpha \beta}{}^\gamma \xi^\beta {\cal A}_{\underline a \gamma} + {\cal A}_{\underline d \alpha} \Gamma^{\underline d}{}_{\underline {a}}  - \frac 1 2 \, \bar{\epsilon}\, \gamma_{\underline{a}}\, \Psi_{\overline {\cal A}}\,  {\cal E}_\alpha{}^{\overline {\cal A}} \ , \label{TransformationASUSY}
\ee
which using the map \eqref{IndexRelation} can be rewritten as
\begin{eqnarray}
 \delta {\cal A}_{\underline a \overline{\cal B C}}= \widehat {\cal L}_\xi {\cal A}_{\underline a \overline{\cal B C}} - {\cal D}_{\underline a} \xi_{\overline{\cal B C}} +  2 {\cal A}_{\underline a \overline{\cal D} [\overline{\cal C}}\, \xi^{\overline{\cal D}}{}_{\overline{\cal B}]} +  {\cal A}_{\underline d \overline{\cal B C}}\, \Gamma^{\underline{d}}{}_{\underline a} + \, \bar{\epsilon}\, \gamma_{\underline{a}} \left[  \frac g 2 \Psi_{\overline {\cal A}}\, {\cal E}_\alpha{}^{\overline {\cal A}} (t^\alpha)_{\overline{\cal B C}}  \right]\ . \label{TransformationASUSY2}
\end{eqnarray}
On the other hand the projected flux transforms as
\bea
\delta {\cal F}_{\underline a \overline {\cal B C}} &=& \widehat {\cal L}_\xi {\cal F}_{\underline a \overline {\cal B C}} - {\cal D}_{\underline a} \Gamma_{\overline{\cal B C}} + 2 {\cal F}_{\underline a  \overline {\cal D} [\overline{\cal C}} \Gamma^{\overline{\cal D}}{}_{\overline{\cal B}]} + {\cal F}_{\underline d  \overline {\cal B C}} \Gamma^{\underline d}{}_{\underline {a}} + \bar \epsilon  \gamma_{\underline a} \Psi_{\overline {\cal BC}} + \nabla_{[\overline {\cal B}} \bar \epsilon \gamma_{\underline a} \Psi_{\overline {\cal C}]} \ , \label{TransformationFProjectedSUSY}
\eea
where we defined
\be
\Psi_{\overline {\cal AB}} = \nabla_{[\overline {\cal A}} \Psi_{\overline {\cal B}]} - \frac 1 2 \omega^{\overline {\cal D}}{}_{\overline {\cal A B}} \, \Psi_{\overline {\cal D}}  = {\cal D}_{[\overline {\cal A}} \Psi_{\overline {\cal B}]} - \frac 1 4 {\cal F}_{[\overline {\cal A} \underline {cd}} \gamma^{\underline {cd}} \Psi_{\overline{\cal B}]} - \frac 1 2 \, {\cal F}_{\overline {\cal A B}}{}^{\overline{\cal C}} \, \Psi_{\overline{\cal C}}\ . \label{gravitinocurvature}
\ee
We will call this the {\it gravitino curvature} with the caveats that (i) it is not fully covariant, as it includes a non-covariant term to render it fully determined, (ii) it represents in fact a curvature for the full $\Psi_{\overline A}$, which includes gauginos as well. So strictly this is a misnomer, but it helps in highlighting the similarity with the identification in \cite{Bergshoeff:1988nn}. Comparing with \eqref{TransformationASUSY2} we see that the last term in \eqref{TransformationFProjectedSUSY} must be canceled. To cure the mismatch, we must  redefine the flux as follows
\be
\mathcal{F}^*_{\underline{a} \bar{\mathcal{B}} \bar{\mathcal{C}}} = \mathcal{F}_{\underline{a} \bar{\mathcal{B}} \bar{\mathcal{C}}} - \frac12 \bar{\Psi}_{\bar{\mathcal{B}}} \gamma_{\underline{a}} \Psi_{\bar{\mathcal{C}}} \ ,
\ee
such that now
\bea
\delta {\cal F}^*_{\underline a \overline {\cal B C}} &=& \widehat {\cal L}_\xi {\cal F}^*_{\underline a \overline {\cal B C}} - {\cal D}_{\underline a} \Gamma_{\overline{\cal B C}} + 2 {\cal F}^*_{\underline a  \overline {\cal D} [\overline{\cal C}} \Gamma^{\overline{\cal D}}{}_{\overline{\cal B}]} + {\cal F}^*_{\underline d  \overline {\cal B C}} \Gamma^{\underline d}{}_{\underline {a}} + \bar \epsilon  \gamma_{\underline a} \Psi_{\overline {\cal BC}}  \ . \label{TransFstar}
\eea

The following identification equates \eqref{TransformationASUSY2} with \eqref{TransFstar}
\be
\boxed{
\begin{split}
\xi_{\overline {\cal A B}}&= -g  \, \xi_\alpha \, (t^\alpha)_{\overline {\cal A B}} \ =\ \Gamma_{\overline {\cal A B}} \ ,\\
{\cal A}_{\underline a \overline {\cal B C}}&= -g  \, {\cal E}_{\alpha \underline a} \, (t^\alpha)_{\overline {\cal B C}} \ =\ {\cal F}^*_{\underline a \overline {\cal B C}}\ , \\
\Psi'_{\overline{\cal A B}} &\equiv \frac g 2 \Psi_{\overline {\cal D}}\,  {\cal E}_\alpha{}^{\overline {\cal D}} (t^\alpha)_{\overline{\cal A B}} = \Psi_{\overline{\cal A B}}\ .
\end{split} \label{LockingSUSY}}
\ee
In particular, the last identification can be solved for the gaugino dof in terms of derivatives of the double generalized frame and gravitino. This is the supersymmetric extension of what we previously called the generalized Bergshoeff-de Roo identification.

This identification is self-consistent because both sides of the last line transform equally. On the one hand, the extended gravitino curvature (\ref{gravitinocurvature}) transforms as
\bea
\delta \Psi_{\overline {\cal AB}} &=& \xi^P \partial_ P\Psi_{\overline {\cal AB}} - 2 \Psi_{\overline {\cal C} [\overline{\cal A}} \Gamma^{\overline {\cal C}}{}_{\overline {\cal B}]} - \frac 1 4 \Gamma_{\underline {ab}} \, \gamma^{\underline{ab}}\, \Psi_{\overline {\cal AB}} + \frac 1 2 \Psi^{\overline{\cal C}}\, {\cal D}_{\overline {\cal C}} \Gamma_{\overline {\cal A B}} \nn \\
&&+\, \frac 1 2 \,{\cal F}^{\underline c}{}_{\overline {\cal A B}} \, {\cal D}_{\underline c} \epsilon - \frac 1 4 \left( {\cal D}_{[{\overline {\cal A}}} {\cal F}_{\overline {\cal B}]\underline{ab}} + {\cal F}_{\underline a\overline{\cal A}}{}^{\underline c}\, {\cal F}_{\underline b \overline {\cal B} \underline {c} } - \frac 1 2 {\cal F}_{\overline {\cal C A B}} \, {\cal F}_{\underline a b}{}^{\overline {\cal C}} \right)  \gamma^{\underline {ab}} \epsilon \ . \label{DeltaPsi}
\eea
On the other, the combination it is identified with in \eqref{LockingSUSY}
\be
\Psi'_{\overline{\cal A B}} = \frac g 2 \, \Psi_{\overline {\cal D}} \, {\cal E}_\alpha{}^{\overline {\cal D}}\, (t^\alpha)_{\overline {\cal A B}}\ ,
\ee
transforms as
\bea
\delta \Psi'_{\overline {\cal AB}} &=& \xi^P \partial_ P\Psi'_{\overline {\cal AB}} - 2 \Psi'_{\overline {\cal C} [\overline{\cal A}} \xi^{\overline {\cal C}}{}_{\overline {\cal B}]} - \frac 1 4 \Gamma_{\underline {ab}} \, \gamma^{\underline{ab}}\, \Psi'_{\overline {\cal AB}} + \frac 1 2 \Psi^{\overline{\cal C}}\, {\cal D}_{\overline {\cal C}} \xi_{\overline {\cal A B}} \nn \\
&&+\, \frac 1 2 \,{\cal A}^{\underline c}{}_{\overline {\cal A B}} \, {\cal D}_{\underline c} \epsilon + \frac 1 4 \left( {\cal D}_{\underline a} {\cal A}_{\underline b \overline{\cal A B}} + {\cal A}_{\underline a\overline{\cal A} }{}^{\overline {\cal C}}\, {\cal A}_{ \underline {b}\overline {\cal B C} } - \frac 1 2 {\cal A}_{\underline c\overline {\cal A B}} \, {\cal F}_{\underline {a b}}{}^{\underline c} \right)  \gamma^{\underline {ab}} \epsilon \ . \label{DeltaPsiPrima}
\eea
In deriving the last expression we used the following identity
\be
{\cal E}_\alpha{}^{\overline {\cal C}} {\cal F}_{\overline {\cal C} \underline{ab}} = - {\cal E}_{\alpha}{}^{ \underline c} {\cal F}_{\underline {cab}} + 2 {\cal D}_{[\underline a} {\cal E}_{\alpha \underline b]} + g f_{\alpha \beta \gamma} {\cal E}^\beta{}_{\underline a} {\cal E}^\gamma{}_{\underline b} \ .
\ee
Employing the identifications (\ref{LockingSUSY}), all the terms in (\ref{DeltaPsi}) and (\ref{DeltaPsiPrima}) can be identified straightforwardly, except for the terms in brackets. It is easy to see however that also those terms coincide exactly using the following projected form of the extended generalized Bianchi identities (\ref{extendedBI})
\be
{\cal D}_{[\overline {\cal A}} {\cal F}_{\overline {\cal B}] \underline {ab}} + {\cal D}_{[\underline a} {\cal F}_{\underline b] \overline{\cal  A B}} + {\cal F}_{[\underline a | \overline {\cal A}}{}^{\cal C} {\cal F}_{|\underline b] \overline {\cal B }{\cal C} } - \frac 1 2 {\cal F}_{\cal C\overline {\cal A B}} {\cal F}_{\underline {ab}}{}^{\cal C} = 0 \ .
\ee
Note that in these equations there is no distinction between ${\cal F}$ and ${\cal F}^*$ because we are working to leading order in fermions only. We would also like to emphasize that the identifications \eqref{LockingSUSY} are exact, and totally independent of the gauge fixing.

A perturbative treatment of the supersymmetric case and the proof that it exactly reproduces the results of \cite{Bergshoeff:1989de} to first order in $\alpha'$ will be presented in  \cite{LNS}.

\section{Outlook}\label{sec::outlook}

We considered the ${\cal N} = 1$ supersymmetric heterotic DFT. The duality group ${\cal G} = O(10,10+k)$ was decomposed in terms of $G=O(10,10)$ multiplets. The physical $G$-covariant dof are a generalized frame and a constrained $G$-vector. We pointed out that the $G$-vector could be identified with certain generalized fluxes provided the heterotic gauge group $\cal K$ were taken to coincide with $\overline {\cal H}= \overline{O(1, 9 + k)}$. A priori this generalized Bergshoeff-de Roo identification seems unlikely to succeed because the dimension of both groups differs for finite $k$. We are then forced to consider infinite-dimensional groups and establish a dictionary between them. A more rigourous treatment of this mathematical structure is lacking, and deserves more attention in the future. The procedure allowed us to lock the $G$-vector in terms of the projected generalized fluxes. This is not a gauge fixing, but a mechanism that actually reduces the physical dof. A similar locking is necessary in the supersymmetric sector, where the gauginos must be locked in terms of a generalized gravitino curvature. The generalized identification is imposed {\it by hand} and it would be nice to find a broader framework from which it arose naturally. Interestingly the identification is exact, so it presumably captures an infinite tower of  $\alpha'$ corrections, giving rise to an exact heterotic generalized Green-Schwarz transformation that closes and preserves the constraints on the fields by construction. We show that a perturbative $\alpha'$ expansion is possible, finding at first order the known heterotic generalized Green-Schwarz transformation of \cite{Marques:2015vua}. We tested the proposal by computing the following ${\cal O}(\alpha'{}^2)$ order, finding a consistent higher derivative completion that was previously unknown.

The results open the door to a large number of questions and future directions. We elaborate on some important points:
\begin{itemize}

\item {\bf Gauge fields.} We got rid of gauge fields by identifying them with generalized fluxes. This was just a procedure implemented to reach the gravitational higher derivative corrections in a duality covariant form. However, the heterotic string and gauged supergravities in general contain gauge fields as proper independent dof (e.g. those of ${\cal K} = SO(32)$ or ${\cal K} = E_8 \times E_8$). Reincorporating these fields is a relatively simple task that can be done in two ways. (i) Starting from a ${\cal G} = O(D,D + k + k')$, one can lock $k$-vectors and leave the other $k'$ free. This was done for instance in \cite{Bedoya:2014pma},\cite{Coimbra:2014qaa},\cite{Lee:2015kba}. (ii) Alternatively, one could consider the $\alpha'$ expansion of the $O(D,D)$ generalized Green-Schwarz transformation, and promote the $O(D,D)$ to an $O(D,D+k')$ thus incorporating $k'$ dynamical vector fields. This was done in \cite{Baron:2017dvb} for generic gauged supergravities, finding in the heterotic case up to quartic powers of the gauge curvatures $F^4$ in exact coincidence with those computed in \cite{Bergshoeff:1989de}.

\item {\bf Quartic Riemann interactions.} Our results provide an all-order supersymmetric duality covariant completion of the Green-Schwarz transformation. A natural question is what kind of interactions are under its reach.

    The standard bosonic Green-Schwarz transformation of the Kalb-Ramond field generates Chern-Simons terms in its three-form curvature. Its first order duality covariant completion \cite{Marques:2015vua} fixes the connection to the heterotic one, and moreover {\it requires and fixes} quadratic Riemann interactions. This is not surprising because T-duality mixes the Kalb-Ramond and the gravitational sectors. Supersymmetry is another ingredient that constrains interactions. It was shown in \cite{Bergshoeff:1989de} that the supersymmetric completion of the Lorentz Chern-Simons terms induced by the Green-Schwarz transformation require deformed supersymmetric transformations that lead to quartic Riemann interactions (which are mirrored to their corresponding gauge field analogs). There is a different set of quartic Riemann terms that have no analog in the gauge sector. Based on the symmetries shared with the construction in \cite{Bergshoeff:1989de}, it is possible that  the framework presented here captures the first set, but not the second set of interactions.

\item {\bf Bi-parametric deformations.} The first order heterotic Green-Schwarz transformation belongs to a bi-parametric family of deformations \cite{Marques:2015vua} (see also \cite{Hohm:2014xsa})
\be
\delta_\Lambda E_M{}^A = E_M{}^B \Lambda_B{}^A + a \, \partial_{[\underline M} \Lambda_{\underline c}{}^{\underline b} \, F_{\overline N] \underline b}{}^{\underline c} \, E^{N A} - b \, \partial_{[\overline M} \Lambda_{\overline c}{}^{\overline b} \, F_{\underline N] \overline b}{}^{\overline c} \, E^{N A} + {\cal O}(\alpha'{}^2) \ ,
\ee
corresponding to the cases $a = 0$ or $b = 0$ (which are the same up to a change of sign of the Kalb-Ramond field), that more generally captures the gauge transformations of the bosonic string \cite{Metsaev:1987zx} ($a=b$) and the HSZ theory \cite{Hohm:2013jaa} ($a = -b$).

First one can ask wether there is an extension of the framework considered here featuring both deformations. Since the two parameters $a$ and $b$ account for the different groups $\underline H$ and $\overline H$ respectively, it seems likely that these deformations will arise from a further extended tangent space ${\cal G} = O(D+k,D+k)$, whose physical dof can be parameterized in terms of a double frame, two pairs of covariantly constrained $O(D,D)$ vectors and also scalars in the coset $\frac {O(k,k)}{O(k)\times O(k)}$. The identifications would then be a little more involved because, although the vector fields are likely to be identifiable with projections of the extended generalized fluxes, the scalar fields would have to be identified as well, with the complication that they are also covariantly constrained to be an element of $O(k,k)$. We plan to study the general case in the future. Of course, yet another question is whether new deformations start beyond the first order.

Notice that in the supersymmetric formulation the fermions are $\underline {\cal H}$ spinors. In $10$ dimensions we would then need $\underline {\cal H} = O(9,1)$ thus forbidding the supersymmetrization of the case $a \neq 0$. ${\cal N} = 1$ Supersymmetry then reduces the space of parameters to a single deformation parameterized by $b$ (for the supersymmetry conventions employed in this paper). This is expected because the bosonic string and the HSZ theory do not admit a supersymmetric completion.

\item {\bf Maximal supersymmetry.} The proper framework to address duality covariant higher derivative corrections in theories with maximal supersymmetry is Exceptional Field Theory. Consider as an example the case of 4 space-time dimensions \cite{Hohm:2013uia} with $E_{7(7)}$ duality symmetry. In order to examine a possible uplift of the deformations considered here to the maximal theory, the natural route  would be to first take this DFT construction to a four-dimensional Kaluza-Klein formulation \cite{Hohm:2013nja} and then explore how the duality group embeds into $E_{7(7)}$. While this is certainly possible when the duality group is $O(6,6)$, the case $O(6,6+k)$ does not admit such an uplift. This makes us believe that the deformations considered here are not consistent with maximal supergravity, nor exceptional symmetries. Perhaps the results in \cite{Blair:2018lbh} shed light on this point.

 Instead, the higher derivatives in maximal theories start at eight derivatives ${\cal O}(\alpha'{}^3)$ through quartic Riemann interactions -which however are different in structure from the heterotic ones-, among others beyond the gravitational sector. A possibility is that generating these corrections would require a new deformation starting at this order, in which case the standard EFT action would not be invariant, and duality covariant eight-derivative terms (and higher) would be necessary.

\item {\bf Extended tangent space approach.} Previous treatments of first-order higher derivatives through finite extensions of the generalized tangent space \cite{Bedoya:2014pma} must be put into scrutiny in light of the results discussed here. The identification performed here generates an infinite extension of the generalized tangent space that accommodates all higher orders. The extended generalized frame ${\cal E}_{\cal M}{}^{\cal A}$ contains extra directions ${\cal E}_\alpha{}^{\cal A}$ and ${\cal E}_{\cal M}{}^{\overline \alpha}$, beyond the double ones ${\cal E}_M{}^A$. The extended directions take values in the adjoint of ${\cal K}$, which after being identified with $\overline {\cal H}$ becomes infinite-dimensional. There is a subtlety though, in that when converting indices $\alpha \to (\overline {\cal A B}) = (\overline {ab}, \overline {a \beta} , \overline{\alpha b}, \overline{\alpha \beta})$, the components $\overline {a b}$ and $\overline {\alpha \beta}$ were shown to start at the same order in perturbations. The $\overline \alpha$ indices are then further identified over and over generating the infinite dimensional extended tangent space -see discussion below equation (\ref{Faalphabeta})-. The point we want to make is that first order corrections are distributed all over the infinite dimensional tangent space, and not only through a single finite extension as in \cite{Bedoya:2014pma}, but through infinite first order replicas. We have seen that all these contributions converge and add up to the expected first order deformation, so both approaches are effectively equivalent to first order: lifting the first order deformations \eqref{deltaEoverline1} and \eqref{deltaEunderline1} to generalized diffeomorphisms in an extended tangent space is a trivial task. However, seeking a lift for the second order deformations to the generalized Green-Schwarz transformations (\ref{Second1}, \ref{Second2}) looks more complicated, casting doubts on further higher derivatives being accounted for through finite extensions of the generalized tangent space.

\item {\bf Action.} We have only discussed exact gauge transformations, but finding the exact gauge invariant action can be done by following the same procedure. One should start from the ${\cal G}$ invariant heterotic ${\cal N} = 1$ supersymmetric DFT action \cite{Hohm:2011ex},\cite{Hohm:2011nu}
\begin{equation}
S_{\mathcal{N}=1} \ = \ \int d^{2D+k} X\,e^{-2d}\left(\mathcal{R}(\mathcal{E},d)
  - \bar{\Psi}^{\bar{\mathcal{A}}}\gamma^{\underline{b}}\nabla_{\underline{b}}\Psi_{\bar{\mathcal{A}}}
 -\bar{\rho}\gamma^{\underline{a}}\nabla_{\underline{a}}\rho-2\bar{\Psi}^{\bar{\mathcal{A}}}\nabla_{\bar{\mathcal{A}}}\rho \right)\ . \label{Action}
\end{equation}
Decomposing the extended generalized frame with respect to $G$-multiplets as in (\ref{FrameParameterizationA}), and performing the identifications (\ref{LockingSUSY}) should lead to the final action. One could then realize a perturbative $\alpha'$ expansion to find the action order by order. The only non trivial step here is that we should get rid of the Greek dummy indices by implementing manipulations similar to those discussed above (\ref{Second1}).

It is possible that apart from  \eqref{Action}, there exist higher derivative invariants that trigger their own tower of $\alpha'$ corrections. They should be invariant under duality symmetries, extended generalized diffeomorphisms, extended Lorentz transformations and supersymmetry. If they exist, the $G$-decomposition, identifications, and derivative expansion would proceed in exactly the same way as here.

\item {\bf Non-perturbative treatment.} Possibly the most important question is how to deal with this deformation exactly, without performing an $\alpha'$ expansion. A similar question has been addressed is the HSZ setup \cite{Hohm:2013jaa}. Both frameworks are similar in that there is a closed non-perturbative form of the gauge transformations and action, which can then be perturbed in a derivative expansion (see \cite{Hohm:2014eba},\cite{Hohm:2015mka},\cite{Naseer:2016izx},\cite{Hohm:2016lim},\cite{Lescano:2016grn}), such that when written in terms of an $O(D,D)$ generalized frame or metric acquires infinite $\alpha'$ corrections.  It is certainly desirable to learn what kind of information can be extracted from these closed and exact expressions.

\item {\bf Solutions.} Exact all order equations of motion (eom) can be derived from \eqref{Action}. One could then explore higher derivative corrections to supergravity solutions (first order deformations were considered in \cite{Edelstein:2018ewc},\cite{Chimento:2018kop}), and even aim at obtaining exact solutions to all orders. Interestingly, one can perform generalized Scherk-Schwarz reductions \cite{Aldazabal:2011nj},\cite{Geissbuhler:2013uka} of the $\alpha'$-DFT action obtained from (\ref{Action}) to generate higher derivatives corrections in gauged supergravity. In \cite{Baron:2017dvb} the first order scalar potential of half-maximal gauged supergravity was computed in the embedding tensor formalism and a moduli stabilization analysis was addressed to first order. The results here could in principle allow for a non-perturbative treatment that could shed light on issues such as moduli stabilization, supersymmetry breaking, etc. in lower dimensional half-maximal gauged supergravities.

    On a slightly different page, using these results as a solution generating technique in supergravity would require the knowledge of finite double Lorentz transformations, as opposed to the infinitesimal ones considered here. This is due to the fact that in supergravity the double Lorentz group  $\underline H \times \overline H$ is broken to a single Lorentz group, and then after generic T-dualities a finite compensating double Lorentz transformation would be required.

\item {\bf Background independence.} It was argued in \cite{Hohm:2016lge} that in order to achieve manifest background independence, a duality symmetric formulation of higher derivative interactions would require gauge degrees of freedom for enhanced gauged symmetries. This is known in the context of first order $\alpha'$ corrections to DFT \cite{Marques:2015vua} which highly rely on the frame formulation for double Lorentz symmetries. A natural question is wether higher derivatives would require further  enhanced gauge symmetries. We see that in the heterotic case the standard double Lorentz symmetries of DFT are already enough to account for all the $\alpha'$ corrections in the universal gravitational sector considered here in a manifestly background independent way.

\end{itemize}

\section*{Acknowledgments} We are in debt to O. Hohm for exceedingly interesting
remarks and comments, and collaboration at early stages of this project. We also thank C. Nu\~nez and A. Rodriguez for sharing their project with us, and C. Hull, D. Waldram and  B. Zwiebach for enlightening discussions. Additionally, we acknowledge F. Hassler and A. Gitsis for comments on the manuscript (see footnote \ref{typo}). WB receives financial support from ANPCyT-FONCyT (PICT-2015-1525). Our work is supported by CONICET.

\section*{Appendix}

\begin{appendix}

\section{Alternative identifications}

We have discussed equivalences between (composite) dof based on the their gauge and supersymmetric transformations. These equivalences were then used to {\it lock} or {\it fix} one set of dof in terms of the other. The procedure is not a gauge fixing, as it reduces the number of physical degrees of freedom rather that eliminating gauge redundancies. In the context of higher derivatives in heterotic string theory, these equivalences go back to \cite{Bergshoeff:1988nn}, where the gauge fields (for the $SO(32)$ or $E_8 \times E_8$) and a specific Lorentz $SO(1,9)$ spin connection (containing torsion proportional to three-form curvature of the Kalb-Ramond field) were shown to transform somewhat symmetrically, something obvious for gauge symmetries but less clear for supersymmetry, in which case also the gauginos must be identified with the gravitino curvature. This equivalence was exploited to compute the first order in $\alpha'$ action, and later expanded to quartic Riemann interactions in \cite{Bergshoeff:1989de} through a Noether procedure. These ideas were employed in a number of recent works, some of which we mention below:

\begin{itemize}
\item In the context of generalized geometry, \cite{Coimbra:2014qaa} considered an extended tangent space decomposed with respect to $GL(D) \in {\cal G}$. The identification was established after the $GL(D)$ decomposition between the one-form components of the generalized frame, and a generalized spin connection compatible with a reduced structure $\overline H \in \overline {\cal H}$. It was argued there that this is only possible if the connection contains a non-vanishing intrinsic torsion.

\item In the context of DFT, \cite{Lee:2015kba} considered an extended tangent space decomposed with respect to $G \in {\cal G}$, as we do. The difference is that there, the $O(D,D)$ vectors are covariantly constrained in a strong sense, namely they are self-orthogonal and also orthogonal to the generalized derivatives as in \cite{Jeon:2011kp}, and the identification is performed {\it after} solving the constraint. So again this approach fails to provide $O(D,D)$ covariant higher derivative corrections.

\item Also in the context of DFT, \cite{Bedoya:2014pma} considered an extended tangent space decomposed with respect to $GL(D) \in {\cal G}$. The approach here is similar to that in \cite{Coimbra:2014qaa}, with the  difference that the identification relates the one-forms to a component of the generalized $\overline H$ spin connection of DFT, which is $O(D,D)$ covariant. So this approach gets closer to the goal of finding $O(D,D)$ covariant higher derivatives, but still fails to achieve the purpose.
\end{itemize}

In this paper we decided to follow a different route: we are interested in a fully $G= O(D,D)$ covariant identification between the $G$-vector that arises from the $G \in {\cal G}$ decomposition of the extended generalized frame, and (derivatives of) the generalized gravitational degrees of freedom. On top of the one we presented in the paper, which is exact to all orders in derivatives, we also considered some other possibilities which turned out to fail in one way or another. For completion we discuss briefly those that looked more promising:
\begin{itemize}

\item The natural possibility, aligned with the three attempts discussed above is to relate ${\cal K}$ with $\overline H \in \overline {\cal H}$ in a $G$-covariant way. As we discussed, performing the gauge fixing ${\cal E}_{\alpha \overline a} = 0$ and $\delta e_\alpha{}^{\overline \alpha} = 0$, leaves the double frame $E_M{}^A$ and the vector ${\cal A}_{\underline a \alpha} = E^{M}_{\underline a} \, {\cal A}_{M \alpha}$ as the unique degrees of freedom. We should then identify  ${\cal A}_{\underline a \alpha}$ with the $\overline H$ components ${\cal F}_{\underline a \overline {bc}}$ of the generalized $\overline {\cal H}$ spin connection (and not the full $\overline {\cal H}$ generalized spin connection as we do in this paper). We would then need a map between $\cal K$ and $\overline { H}$  given by the Lorentz generators $(t_\alpha)_{\overline {ab}}$
    \be
    [t_\alpha ,\, t_\beta] = f_{\alpha \beta}{}^\gamma t_\gamma \ , \ \ \ \  (t^\alpha)_{\overline {a b}} (t_\beta)^{\overline {ab}} = X_R \, \delta^\alpha_\beta \ , \ \ \ \ (t^\alpha)_{\overline {a b}} (t_\alpha)^{\overline {c d}} = X_R \, \delta^{\overline{cd}}_{\overline {ab}} ,
    \ee

    such that indices can be converted from one group to the other
    \bea
          V_{\overline {ab}} = - g\, V^\alpha\, (t_\alpha)_{\overline {a b}} \ .
    \eea
    The gauge transformations of both quantities are
    \bea
    \delta {\cal A}_{\underline a \overline{b c}} &=& \xi^P \partial_P {\cal A}_{\underline a \overline {bc}} + \Gamma^{\underline d}{}_{\underline a} \, {\cal A}_{\underline d \overline { b c}} - {\cal D}_{\underline a} \xi_{\overline {b c}} + 2 {\cal A}_{\underline a \overline d [\overline c}  \xi^{\overline d}{}_{\overline b]}\ , \label{transfA}\\
    \delta {\cal F}_{\underline a \overline{b c}} &=& \xi^P \partial_P {\cal F}_{\underline a \overline {bc}} + \Gamma^{\underline d}{}_{\underline a} \, {\cal F}_{\underline d \overline { b c}} - {\cal D}_{\underline a} \Gamma_{\overline {b c}} + 2 {\cal F}_{\underline a \overline d [\overline c}  \Gamma^{\overline d}{}_{\overline b]}+ 2 {\cal F}_{\underline a \overline \delta [\overline c}  \Gamma^{\overline \delta}{}_{\overline b]}\ . \label{transfF}
    \eea
    There is a clear mapping between these transformations, except for the last term in (\ref{transfF}), which cannot be set to zero because the gauge fixing locks the off-diagonal component of the $\overline {\cal H}$ parameter to a non-vanishing value. An identification is then not possible in   general. However, after taking the parameter to its gauge fixed value (\ref{GaugedFixedParameters}) it can be seen that the last term in (\ref{transfF}) starts from one order higher in derivatives than the rest, and so this approach is perfectly consistent to first order in $\alpha'$, in which case the extended fluxes can be replaced by double fluxes $F_{\underline a \overline {bc}}$ without any loss of generality. It is easy to check that this identification reproduces the first order generalized Green-Schwarz transformation, and also gives rise to the extended tangent space approach of \cite{Bedoya:2014pma} after a $GL(D)$ decomposition.

\item One could try to avoid the gauge fixing, so as to have freedom to set $\Gamma_{\overline{a \alpha}} = 0$, thus solving the problem of the previous attempt. If so, one should now also identify the degrees of freedom ${\cal A}_{\overline a \alpha} = {\cal E}_{\alpha \overline a}$ which can no longer be set to zero. Using the relation between $\cal K$ and $\overline H$ above we find
    \bea
    \delta {\cal A}_{A \overline{b c}} &=& \xi^P \partial_P {\cal A}_{A \overline {bc}} + \Gamma^{D}{}_{A} \, {\cal A}_{D \overline { b c}} - {\cal D}_{A} \xi_{\overline {b c}}+ 2 {\cal A}_{A \overline d [\overline c}  \xi^{\overline d}{}_{\overline b]} \ , \label{transffullA}
    \eea
    which is exactly the way in which the $\overline H$ components of the generalized spin connection transform (an extended tangent space containing the generalized spin connection was also considered in \cite{Polacek:2013nla})
    \bea
    \delta \omega_{A \overline{b c}} &=& \xi^P \partial_P \omega_{A \overline {bc}} + \Gamma^{D}{}_{A} \, \omega_{D \overline { b c}} - {\cal D}_{A} \Gamma_{\overline {b c}} + 2 \omega_{A \overline d [\overline c}  \Gamma^{\overline d}{}_{\overline b]}\ . \label{transfomega}
    \eea
    Identifying these degrees of freedom wouldn't really solve the problem then, because only some projections like $\omega_{[\overline {a b c}]} = \frac 1 3 {\cal F}_{\overline {a b c}}$ are determined, and the transformation of the double frame would depend on undetermined components.

\item Independently of what one identifies ${\cal E}_{\alpha \overline a}$ with, a different question is wether the dependence on this component can be eliminated from the gauge transformations through field-redefinitions. Assuming ${\cal E}_{\alpha A}$ is of first order in $\alpha'$, the leading order transformations of the double frame and ${\cal E}_{\alpha A}$ are given by
    \bea
    \delta E_M{}^A &=& \widehat {\cal L}_\xi E_M{}^A + E_M{}^B\, \Gamma_B{}^A  - E_{M B}\, D^{[A}\xi^\alpha\, {\cal E}_\alpha{}^{B]} + {\cal O}(\alpha'{}^2) \ , \\
    \delta {\cal E}_{\alpha A} &=& \widehat {\cal L}_\xi {\cal E}_{\alpha A} - D_A \xi_\alpha + g f_{\alpha \beta}{}^\gamma \xi^\beta {\cal E}_{\gamma A} + {\cal E}_{\alpha B} \Gamma^B{}_A + {\cal O}(\alpha') \ .
    \eea
    If we now redefine the Lorentz parameters to first order
    \bea
    \Gamma^{\underline{ab}} &=& \Lambda^{\underline {ab}} + D^{[\underline a} \xi^\alpha \, {\cal E}_\alpha{}^{\underline b]} \ , \\
    \Gamma^{\overline{ab}} &=& \Lambda^{\overline {ab}} + D^{[\overline a} \xi^\alpha \, {\cal E}_\alpha{}^{\overline b]} \ ,
    \eea
    the transformations of the double frame become
    \bea
    \delta E_M{}^{\overline a} &=& \widehat {\cal L}_\xi E_M{}^{\overline a} + E_M{}^{\overline b} \Lambda_{\overline b}{}^{\overline a} - \frac 1 2 E_{M \underline b}\, D^{\overline a} \xi^\alpha \, {\cal E}_\alpha{}^{\underline b}  + \frac 1 2 E_{M \underline b}\, D^{\underline b} \xi^\alpha\, {\cal E}_\alpha{}^{\overline a} + {\cal O}(\alpha'{}^2) \ ,\\
    \delta E_M{}^{\underline a} &=& \widehat {\cal L}_\xi E_M{}^{\underline a} + E_M{}^{\underline b} \Lambda_{\underline b}{}^{\underline a} - \frac 1 2 E_{M \overline b}\, D^{\underline a} \xi^\alpha \, {\cal E}_\alpha{}^{\overline b}  + \frac 1 2 E_{M \overline b}\, D^{\overline b} \xi^\alpha\, {\cal E}_\alpha{}^{\underline a} + {\cal O}(\alpha'{}^2) \ . \quad
    \eea
    Our purpose is to eliminate ${\cal E}_{\alpha \overline a}$ through redefinitions. Redefining the double frame as follows
    \bea
    \widetilde E_M{}^{\overline a} &=& E_M{}^{\overline a} + \frac 1 2 {\cal E}_\alpha{}^{\overline a}\, {\cal E}^{\alpha \underline b} \, E_{M \underline b} \ ,\\
    \widetilde E_M{}^{\underline a} &=& E_M{}^{\underline a} - \frac 1 2 {\cal E}_\alpha{}^{\underline a}\, {\cal E}^{\alpha \overline b} \, E_{M \overline b} \ ,
    \eea
    achieves the purpose
    \bea
    \delta \widetilde E_M{}^{\overline a} &=& \widehat {\cal L}_\xi \widetilde E_M{}^{\overline a} + \widetilde E_M{}^{\overline b} \Lambda_{\overline b}{}^{\overline a} - \widetilde E_{M\underline b} \, D^{\overline a} \xi^\alpha \, {\cal E}_\alpha{}^{\underline b} + {\cal O}(\alpha'{}^2) \ , \\
    \delta \widetilde E_M{}^{\underline a} &=& \widehat {\cal L}_\xi \widetilde E_M{}^{\underline a} + \widetilde E_M{}^{\underline b} \Lambda_{\underline b}{}^{\underline a} + \widetilde  E_{M\overline b} \, D^{\overline b} \xi^\alpha \, {\cal E}_\alpha{}^{\underline a} + {\cal O}(\alpha'{}^2) \ .
    \eea
    We tried to pursue this procedure to the next order, but the treatment becomes cumbersome, and then one wonders if the effort is worth considering we already have an exact identification that can be treated easily and expanded perturbatively order by order.

\end{itemize}
\end{appendix}

\end{document}